\documentclass[article]{aastex61}

\newcommand{\water}{H$_2$O}
\newcommand{\SPHEREx}{{\sl SPHEREx}}
\newcommand{\numsources}{$8.6\times10^6$}

\newcommand{\acronym}{SPLICES}
\newcommand{\shortcatalog}{Target List}
\newcommand{\version}{version 7.1}

\begin{document}
\title{The SPHEREx Target List of Ice Sources (SPLICES)}
\shorttitle{SPHEREx Ice Target List}
\shortauthors{Ashby et al.}

\author[0000-0002-3993-0745]{Matthew L.\ N.\ Ashby}
\affiliation{Center for Astrophysics  $|$ Harvard \& Smithsonian, 
Optical and Infrared Astronomy Division, Cambridge, MA 01238, USA}

\author[0000-0002-5599-4650]{Joseph L.\ Hora}
\affiliation{Center for Astrophysics  $|$ Harvard \& Smithsonian, 
Optical and Infrared Astronomy Division, Cambridge, MA 01238, USA}

\author[0000-0002-3889-0630]{Kiran Lakshmipathaiah}
\affiliation{Indian Institute of Space science and Technology, 
Thiruvananthapuram 695547, Kerala, India}

\author[0000-0002-3477-6021]{Sarita Vig}
\affiliation{Indian Institute of Space science and Technology, 
Thiruvananthapuram 695547, Kerala, India}

\author[0000-0001-5021-0071]{Rama Krishna Sai Subrahmanyam Gorthi}
\affiliation{Department of Electrical Engineering, Indian Institute of Tirupati, Andra Pradesh, 517619, India}

\author[0000-0002-5016-050X]{Miju Kang}
\affiliation{KASI (Korea Astronomy and Space Science Institute), 776 Daedeok-daero, Yuseong-gu, Daejeon 34055, Republic of Korea}

\author[0000-0003-1841-2241]{Volker Tolls}
\affiliation{Center for Astrophysics  $|$ Harvard \& Smithsonian, 
Optical and Infrared Astronomy Division, Cambridge, MA 01238, USA}

\author[0000-0002-6025-0680]{Gary J.\ Melnick}
\affiliation{Center for Astrophysics  $|$ Harvard \& Smithsonian, 
Optical and Infrared Astronomy Division, Cambridge, MA 01238, USA}

\author[0000-0003-4990-189X]{Michael W.\ Werner}
\affiliation{Jet Propulsion Laboratory, California Institute of Technology, 4800 Oak Grove Drive, Pasadena, CA 91109, USA}

\author[0000-0002-4650-8518]{Brendan P.\ Crill}
\affiliation{Jet Propulsion Laboratory, California Institute of Technology, 4800 Oak Grove Drive, Pasadena, CA 91109, USA}
\affiliation{California Institute of Technology, 1200 E. California Boulevard, Pasadena, CA 91125, USA}

\author[0000-0001-5382-6138]{Daniel C.\ Masters}
\affiliation{Infrared Processing and Analysis Center, 
California Institute of Technology, Pasadena, CA 91125, USA}

\author[0000-0003-1894-1880]{Carlos Contreras Pe\~{n}a}
\affiliation{Department of Physics and Astronomy, Seoul National University, Seoul 08826, Republic of Korea}

\author[0000-0003-3119-2087]{Jeong-Eun Lee}
\affiliation{Department of Physics and Astronomy, Seoul National University, Seoul 08826, Republic of Korea}

\author[0000-0001-8064-2801]{Jaeyeong Kim}
\affiliation{KASI (Korea Astronomy and Space Science Institute), 776 Daedeok-daero, Yuseong-gu, Daejeon 34055, Republic of Korea}

\author[0000-0002-3808-7143]{Ho-Gyu Lee}
\affiliation{KASI (Korea Astronomy and Space Science Institute), 776 Daedeok-daero, Yuseong-gu, Daejeon 34055, Republic of Korea}

\author[0000-0001-6216-0462]{Sung-Yong Yoon}
\affiliation{School of Space Research, Kyung Hee University, 1732, Deogyeong-daero, Giheung-gu, Yongin-si, Gyeonggi-do 17104, Republic of Korea }
\affiliation{KASI (Korea Astronomy and Space Science Institute), 776 Daedeok-daero, Yuseong-gu, Daejeon 34055, Republic of Korea}

\author[0000-0001-9842-639X]{Soung-Chul Yang}
\affiliation{KASI (Korea Astronomy and Space Science Institute), 776 Daedeok-daero, Yuseong-gu, Daejeon 34055, Republic of Korea}

\author[0000-0002-8763-1555]{Nicholas Flagey}
\affiliation{Space Telescope Science Institute, 3700 San Martin Dr., Baltimore, MD, 21218, USA}

\author[0000-0003-4205-4800]{Bertrand Mennesson}
\affiliation{Jet Propulsion Laboratory, MS 264-767, 4800 Oak Grove Drive,
Pasadena, CA 91109}
\begin{abstract}

One of the primary objectives of the \SPHEREx\ mission is to understand
the origin of molecules such as \water, CO$_2$, and other volatile compounds at
the early stages of planetary system formation.  Because the vast majority of these
compounds -- typically exceeding 95\% -- exist in the solid phase rather than the gaseous
phase in the systems of concern here, the observing strategy planned
to characterize them is
slightly unusual.  Specifically, \SPHEREx\ will target highly obscured sources throughout 
the Milky Way, and observe the species of concern in absorption against background 
illumination.  \SPHEREx\ spectrophotometry will yield ice column density measurements for 
 millions of obscured Milky Way sources of all ages
and types.  By correlating those column densities with source ages, 
the \SPHEREx\ mission will shed light on whether those molecules were formed
{\sl in situ} along with their nascent stellar systems, or whether instead they formed elsewhere
and were introduced into those systems after their formation.  To that end, this work describes
 \version\ of the \SPHEREx\ Target {\sl L}ist of {\sl Ice} {\sl S}ources (SPLICES) for the community.  
 It contains \numsources\ objects brighter than $W2\sim12$\,Vega mag over much of the sky, 
 principally within a broad strip running the length of the Milky Way midplane, 
but also within high-latitude molecular clouds and even the Magellanic Clouds.

\end{abstract}


\section{Introduction} \label{sec:intro}

Using spectra obtained with the {\sl Infrared Space Observatory} (ISO)'s Short Wavelength Spectrometer,
the groundbreaking \citet{2004ApJS..151...35G} investigation first revealed the rich variety 
of infrared ice absorption features toward obscured Milky Way sources.  
Those features include a typically strong and broad water ice line at 3.0\,$\mu$m, as well as
a number of narrower ice features due to such species as CO, CO$_2$, $^{13}$CO$_2$, and 
XCN (e.g., OCN$^-$; \citealt{2004ApJS..151...35G,2005A&A...441..249V,2022ApJS..259...30M})
which only became accessible
once a sufficiently sensitive spectrometer was launched into orbit above Earth's obscuring atmosphere.  
Presently, only a few of the brightest such objects
have been observed spectroscopically from the ground, yielding evidence for water ice absorption \citep{1998AJ....116..868I}.
In combination with more recent {\sl AKARI} ice spectra toward young stellar 
objects \citep{2022ApJ...935..137K,2012A&A...538A..57A} and molecular clouds \citep{2013ApJ...775...85N}, a total of roughly 200 relatively bright Milky Way targets have been observed from space-based observatories with sufficient sensitivity to characterize the column densities of ices along the lines of sight.  

Ice spectra such as those returned by {\sl ISO} and {\sl AKARI} are extremely useful for elucidating basic questions about the chemical evolution of stellar and planetary systems, because the vast majority of many volatile species are locked up in the solid phase, as opposed to the gas phase (\citealt{2009ApJ...690.1497H,2015ARA&A..53..541B}, but see also \citealt{2017MNRAS.467.4753N}).  
The \SPHEREx\ mission \citep{2018SPIE10698E..1UK, Crill.2020} will return millions of ice spectra, increasing our knowledge of the abundance of important volatile species throughout the Milky Way by several orders of magnitude.

\SPHEREx\ is a NASA Medium Explorer that will
carry out an all-sky spectroscopic survey from 0.75--5.0\,$\mu$m four times during its baseline mission.  \SPHEREx\
is currently in Phase C of development and is progressing to a planned launch in 2025 February.  
From low Earth orbit,
 \SPHEREx\ will survey the entire sky spectroscopically with a 20\,cm diameter telescope illuminating a focal plane 
populated with six H2RG detector arrays, each configured
with a linear variable filter.  Together the six arrays and filters will cover the 0.75--5.0\,$\mu$m wavelength range with resolving powers that vary from
$R\sim40-130$.  For more details, see \citet{Crill.2020}.

The \SPHEREx\ Ices investigation, one of the three main science programs planned by the \SPHEREx\ science team, offers an unprecedented opportunity to correlate the volatile ices content of stellar systems with their ages with tremendous statistical power.  The millions of \SPHEREx\ spectra will make it possible to determine whether the volatile ice content
 of stellar systems is intrinsic, i.e., whether their ices content is present at the time of formation.  If \SPHEREx\ reveals
 that such a correlation is not present, then another mechanism must account for volatile production, such as their formation 
 outside those systems and subsequent introduction into them.

Because \SPHEREx\ is an all-sky survey mission with a mandate to serve the astronomy community, the \SPHEREx\ team will create and maintain a substantial spectral archive for public access.  However, the \SPHEREx\ pipeline will automatically generate spectra only for targets at predefined locations.  Therefore, a list of targets likely to exhibit ice absorption features has been developed prior to launch, to provide spectra for a wide variety of objects and in sufficient numbers to trace the history of organic molecules from their time of formation within molecular clouds to their incorporation in planetary systems.  The present work describes that list and the criteria used to assemble it.  It is
called the \SPHEREx\ target List of ICE Sources (\acronym).   

This work is structured as follows.  
In Sec.~\ref{sec:selection} we describe how targets were identified for the \SPHEREx\ Ices Investigation.  
Sec.~\ref{sec:xmatching} describes broadband photometry added to the 
catalog based on positional cross-matching, while
Sec.~\ref{sec:adhoc} describes additional science sources included in SPLICES.
Sec.~\ref{sec:validation} describes how the target selection techniques have been validated.
We define the catalog content in Sec.~\ref{sec:content} and illustrate some of its idiosyncracies and limitations in Sec.~\ref{sec:illustrations}.  Finally, we provide some concluding remarks in Sec.~\ref{sec:conclusion}.

\section{Target Selection Criteria} \label{sec:selection}

As noted in many other places (see \citet{2015ARA&A..53..541B} and references therein), 
ice species are typically seen in absorption when a bright background source is viewed through 
cold material along the line of sight.  Indeed, substantial optical depths and the attendant shielding from UV radiation 
are thought necessary for ices to form on dust grains \citep{2009ApJ...690.1497H}.  Thus, when $A_V$ ranges beyond a
threshold level of 1.6\,mag, the ice phase begins to enter the mass budgets for volatile species such as H$_2$O, CO, CO$_2$, and $^{13}$CO$_2$, i.e., those to which \SPHEREx\ is sensitive.  Several magnitudes beyond this threshold, the ice phase is expected to dominate the mass budgets.

The approach adopted here is to assemble candidate ice-bearing sources for the \SPHEREx\ mission by 
identifying bright relatively isolated sources that, based on their red broadband colors, show evidence for significant extinction.   The criteria are defined quantitatively below.

\subsection{Infrared Color Criteria} \label{ssec:color_criteria}

Normal stars, and those in the red clump and red giant branch, 
have characteristically neutral near-infrared (NIR) colors in the absence of extinction.  
For example, Fig.\,1 of \citet{2011ApJ...739...25M} shows 
that such stars in the Milky Way have intrinsic colors $H-W2<1$ \,mag over a 
wide range of stellar temperatures and ages, where the $H$-band magnitude is 
taken from the 2MASS Point Source Catalog (PSC; \citealt{2006AJ....131.1163S}) and 
$W2$ is the 4.6\,$\mu$m magnitude from AllWISE \citep{2014yCat.2328....0C}.  

Here we follow \citet{2011ApJ...739...25M} and take $A_K = 0.918(H - [4.5] - 0.08)$ but
substitute $WISE$ $W2$ magnitudes for the IRAC 4.5\,$\mu$m flux densities.  Adopting a
threshold of $A_V>2$ for the onset of ice formation and using $A_K/A_V = 0.112$ \citep{1985ApJ...288..618R}
yields $H-W2>0.324$\,mag as our first selection criterion. 
The $H-W2$ color criterion was chosen 
as a proxy for $A_V>2$ for main sequence background stars, ensuring sufficient UV shielding for ices to form
if the extinction arises primarily within a single cloud along the line of sight.

Some promising sources will inevitably be missed by this simple criterion.  
Deeply embedded sources in particular might go undetected by the 2MASS 
PSC $H$-band, yet nonetheless be well-detected at $K_s$ where extinction is lower.  
Such $K_s$-detected objects may be a useful addition to \acronym\ because 
they likely represent younger sources that have not had 
sufficient time to emerge from their natal clouds.   
To include them, we augmented \acronym\ 
with a second color selection implemented it at $K_s$, by including $H$-undetected objects which satisfy $K_s - W2 > 0.55$\,mag.  

Reliable application of any color criteria implicitly assume reliable cross-matching between catalogs.  In this work we adopt the associations already established by the {\sl Gaia} team \citep{2022arXiv220800211G} between {\sl Gaia} DR3 and 2MASS PSC source positions based on proximity.  For the cross-matches from 2MASS to {\sl WISE}, we adopt the counterparts listed in AllWISE \citep{2014yCat.2328....0C}.  
We used AllWISE as our mid-IR photometric 
reference instead of the more sensitive unWISE  \citep{2018ApJS..234...39S}   
because bright sources ($W2<8$\,Vega mag) suffer from systematic photometric 
biases in unWISE, which would
significantly complicate validation of \acronym\ using known sources.
In this way we constructed consistent point-source photometry from optical to mid-IR wavelengths for all SPLICES targets.

We carried out the 2MASS-to-{\sl WISE} color selections using the AllWISE catalog 
hosted at the Infrared Processing and Analysis Center.  AllWISE 
implements a 3\arcsec\ matching radius to identify 2MASS PSC counterparts to {\sl WISE}-detected
sources.  Specifically, the 2MASS PSC source that is closest to the position of an AllWISE source (within 3\arcsec) is associated with that source.  AllWISE sources not having exactly one 2MASS source within 3\arcsec\ were rejected in the search using the available {\tt n\_2mass} value in the AllWISE database (see Sec.~\ref{ssec:isolation} for details regarding how position-based searches using the 2MASS coordinates were used to eliminate potentially confused targets with neighbors $<6\farcs2$).  

In addition to the above, sources were required to have a 2MASS PSC $H$-band photometric quality flag {\tt ph\_qual} of either A or B, meaning that they were detected with SNR$\ge10$ or SNR$\ge5$, respectively.  
Candidate deeply embedded sources 
were required to have a 2MASS PSC $K_s$-band photometric quality flag {\tt ph\_qual} of A (meaning that the source was detected 
with SNR$\ge10$ at $K_s$), and any $H$-band photometric quality flag {\sl except} A or B, so as to exclude objects that were rejected on the basis of their $H-W2$ color.

\subsection{Isolation Criterion}\label{ssec:isolation}

Compared to 2MASS or {\sl Spitzer}, \SPHEREx\ has relatively coarse spatial resolution.
The \SPHEREx\ PSFs' FWHMs range from about 2\arcsec\ at 0.75\,$\mu$m 
up to almost 5\arcsec\ at 5\,$\mu$m \citep{Crill.2020}.
Its pixels are 6\farcs2 on a side, significantly larger than those of either 2MASS or {\sl WISE} \citep{2010AJ....140.1868W}.    
Thus, multiple sources in the relatively broad 
\SPHEREx\ beam could potentially contribute to the spectra, especially toward the relatively  
confused Milky Way midplane, contaminating or diluting the desired ice absorption features. 
 
 We used two methods to reject sources with apparent neighbors, i.e., sources that would produce blended {\sl SPHEREx} spectra that would be difficult to interpret.  The first method was to use the AllWISE {\tt n\_2mass} value as described in the previous section to eliminate targets having nearby 2MASS-detected neighbors. 
 The second method was to perform a cross-match with the 2MASS PSC and remove sources with 2MASS neighbors within
 6\farcs2, i.e., separated by less than 1 \SPHEREx\ pixel from their nearest neighbor.  This was implemented using the coordinates of the 2MASS PSC $H$-band detections 
 except for the 
 deeply embedded sources, for which the coordinates correspond to the positions as seen at $K_s$.
 
To validate the 6\farcs2 isolation criterion, we simulated \SPHEREx\ photometry within fields containing multiple sources using the prototype \SPHEREx\ pipeline code, which is based on the Tractor software package 
\citep{2016ascl.soft04008L}.
As inputs, the Tractor requires source positions and the instrument PSF (including the effects of spacecraft pointing jitter).  The SPHEREx pipeline will use positional prior information on the sources to produce a generative model of the scene that is optimized to determine the fluxes (``forced photometry"). The assumption in these simulations is that the positions of the sources on the SPHEREx array are well-known (to $\lesssim0\farcs2$) from the external catalogs.
In a first set of simulations, two sources were placed at varying separations ranging from 0.8 to 3 \SPHEREx\ pixels, and at relative secondary/primary source flux ratios from 0.01 to 1.00.  The primary source was held constant at 12\,AB mag (i.e., it was a bright source detected at very high significance), and sky noise appropriate to SPHEREx Band 6 (the longest-wavelength array; \citealt{Crill.2020}) was added to the frames.  The photometry of the simulated images showed that the primary source could be accurately extracted at all separations down to 1 \SPHEREx\ pixel without increasing the photometric uncertainties above 1\%.  The photometric
uncertainties estimated for the secondary source were $<1$\% up to a flux ratio of at least 1.5, beyond which the uncertainties grow larger than 2\% for the closest separations.  

A second set of simulations was then performed to test the effects of pixel phase on the photometry.  For this second set of tests, the primary source was stepped across a pixel in 1/5 pixel steps, while the secondary source was placed at random positions and angular offsets from the primary.  The resulting simulated Tractor photometry revealed no increase in photometric uncertainty over the range of tested pixel phases and confirmed the results of the first test. 

SPLICES targets with nearby companions $>1.5\times$ as bright as themselves are relatively uncommon, comprising less than 5\% of sources with neighbors at separations $<3$ \SPHEREx\ pixels.  Our ability to extract these sources will ultimately depend on the actual in-flight performance of the instrument and fidelity with which the PSF profile can be reconstructed, which will affect the minimum distance and maximum flux ratio of the sources where our photometric accuracy requirements will be met.  Therefore, rather than exclude such objects from SPLICES entirely, we choose to retain and flag them (specifically, to flag targets having brighter neighbors $>1.5\times$ as bright as themselves in the $H$-band, or $K_s$-band for $K_s$-selected targets) as possible confused sources, and will evaluate them during the mission to determine the relationship of distance and relative flux where the photometry will meet our requirements for ice feature extraction.

\subsection{Regions Included in \acronym} \label{sec:regions}

\begin{figure*}
\begin{center}
\includegraphics[width=1.0\columnwidth]{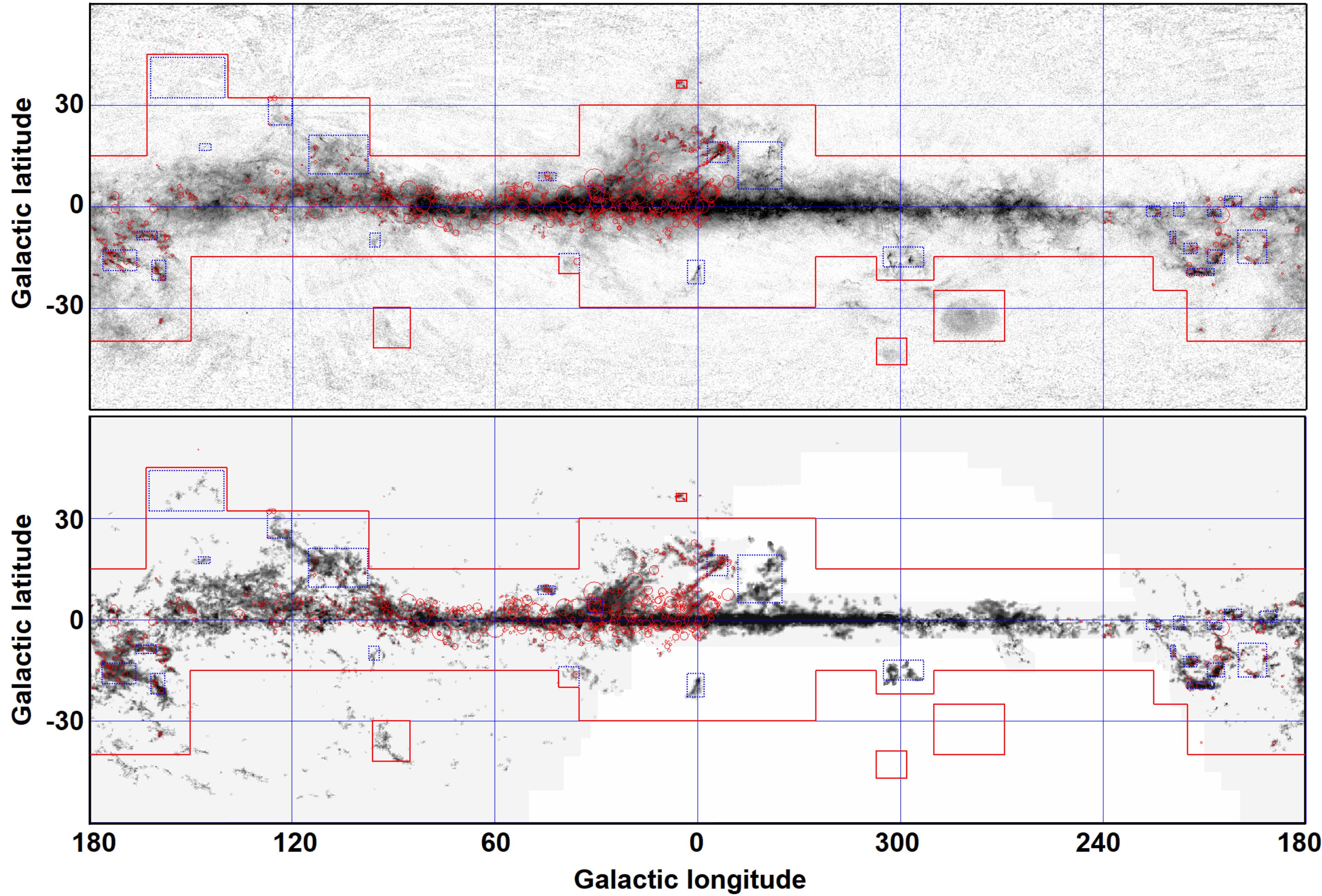}
\end{center}
\caption{Regions surveyed to identify sources likely to present spectra with detectable ice features of H$_2$O, CO, CO$_2$, $^{13}$CO$_2$, and other species in the $3-5$\,$\mu$m wavelength range where \SPHEREx\ is sensitive.  {\sl Upper panel:} The regions are projected onto a background greyscale image representing the 
\citet{2009MNRAS.395.1640R} 
 extinction map stretched from $A_V=0$ (white) to 2\,mag (black).  Solid red lines enclose the fields from which candidate
\SPHEREx\ targets have been drawn.  Dotted blue lines indicate the boundaries of nearby molecular clouds as defined by \citet{2019ApJ...879..125Z}.
Red circles indicate the locations and sizes of `dark nebulae' identified on the Palomar plates by \citet{1962ApJS....7....1L}.
{\sl Lower panel:} same symbols as the upper panel, but the background greyscale image is
the integrated $^{12}$CO(1-0) line intensity survey executed with the CfA 1.2\,m telescope and its southern hemisphere counterpart (e.g., \citealt{2001ApJ...547..792D}; \citealt{2015ARA&A..53..583H}).   The inverted greyscale stretch ranges from $-1$ to 25\,K\,km\,s$^{-1}$, to indicate the extensive CO survey footprint.  The white background indicates regions unobserved by the CfA CO survey except for the Magellanic Clouds (the SMC and LMC maps are described respectively in \citet{1991ApJ...368..173R} and  \citet{1988ApJ...331L..95C}).
}
\label{fig:regions}
\end{figure*}

In this Section we describe the regions surveyed to identify \SPHEREx\ 
ices targets for the current version of \acronym\ (\version).  The regions were chosen to have strong
indications of extinction likely to produce foreground dust attenuation of background stars and YSOs, 
from which spectra the ice features of interest might be detected.  
Their boundaries are defined in Table~\ref{tab:fields} and illustrated in Fig.~\ref{fig:regions}.

{
\tabletypesize{\small}
\setlength{\tabcolsep}{12pt}
\begin{deluxetable*}{lrrrrc}
\tablecaption{\label{tab:fields} Regions Surveyed for Obscured Sources by \SPHEREx}
\tablehead{\colhead{(1)}    & \colhead{(2)}         & \colhead{(3)}         & \colhead{(4)}         & \colhead{(5)}         & \colhead{(6)} \\
           \colhead{Region} & \colhead{$l_{lower}$} & \colhead{$l_{upper}$} & \colhead{$b_{lower}$} & \colhead{$b_{upper}$} & \colhead{Reference} \\ 
           \colhead{}       & \colhead{(deg)   }    & \colhead{(deg)   }    & \colhead{(deg)   }    & \colhead{(deg)   }    & \colhead{}}
\startdata
Milky Way midplane          &   0.0 & 360.0 &  $-$15.0 &    15.0 & \nodata \\
Milky Way central           & $-$35.0 &  35.0 &  $-$30.0 &    30.0 & \nodata \\
Libra                       &   3.2 &   6.2 &     35.0 &   37.25 & \nodata \\
Aquila South                &  35.0 &  41.0 &  $-$20.0 & $-$15.0 & 1       \\
Pegasus                     &  85.0 &  96.0 &  $-$42.0 & $-$30.0 & 1       \\
Cassiopeia/Cepheus Extended &  97.0 & 162.0 &     15.0 &    32.0 & \nodata    \\
Ursa Major/Camelopardalis     & 140.0 & 162.0 &     32.0 &    44.0 & 1    \\
Ophiuchus/California/Taurus & 150.0 & 180.0 &  $-$40.0 &  $-$15.0 & \nodata \\
Monoceros/Orion/Crossbones  & 180.0 & 215.0 &  $-$40.0 &  $-$15.0 & \nodata \\
Canis Major                 & 215.0 & 225.0 &  $-$25.0 &  $-$15.0 & \nodata \\
Chamaeleon Extended         & 290.0 & 307.0 &  $-$22.0 &  $-$15.0 & \nodata \\
Large Magellanic Cloud      & 269.0   & 290.0 & $-$40.0 & $-$25.0 & 2    \\
Small Magellanic Cloud      & 298.0   & 307.0 & $-$47.0 & $-$39.0 & \nodata    \\
\enddata
\tablecomments{Regions surveyed for obscured sources identified 
based on AllWISE and 2MASS PSC colors.  All sources satisfying 
the selection criteria described in the main text and lying on or 
within 
the boundaries listed here are included in the \SPHEREx\ catalog.  }
\tablenotetext{1}{Boundaries defined by \citet{2019ApJ...879..125Z}.}
\tablenotetext{2}{Boundaries chosen to encompass the LMC as traced in the $J$-band
extinction computed by NICEST (Juvela \& Montillaud 2016).}
\end{deluxetable*}}

To include a majority of obscured sources in the \shortcatalog, the Milky Way midplane is surveyed.  
Specifically, the catalog includes all sources satisfying the 
color and isolation criteria described in Sec.~\ref{sec:selection} within $b= \pm15^\circ$ over the entire
midplane ($\ell=0 - 360^\circ$).  Over the central region ($\ell=0\pm35^\circ$), the latitude contraint was widened to include a wider area, $b= \pm30^\circ$.  Thus, SPLICES encompasses the Milky Way $grizY$ coverage provided by the DECam Plane Survey (DEC-PS; 
\citealt{2018ApJS..234...39S,2022arXiv220611909S}) and Pan-STARRs\,1 \citep{2016arXiv161205560C}; our 
aim is to eventually combine SPLICES with optical and near-infrared photometry from DEC-PS and Pan-STARRs\,1 for as many of the cataloged sources as possible
 to aid in source classification and interpretation.
 With this approach we hope to encompass the promising regions of likely ice targets without extending the covered
areas pointlessly to deep-sky fields, where heavy contamination of our source lists by active galactic nuclei would likely result.

Of course not all Milky Way sources lie close to the midplane.  To account for the relatively nearby 
obscured sources that tend to lie at relatively high Galactic latitudes, the coverage was expanded beyond the midplane in several ways.  To begin with, we included the nearby molecular clouds listed in Table\,1 of \citet{2019ApJ...879..125Z}, because of the high likelihood of these regions containing relatively bright targets owing to their proximity.  In many instances the \citet{2019ApJ...879..125Z} cloud boundaries were simply
adopted without modification.  
For others, such as rich regions containing multiple clouds, the \citet{2019ApJ...879..125Z} boundaries were enlarged to encompass all the surrounding 
nebulosity.   
For example, to incorporate  Perseus, Taurus, Orion, and Monoceros, as well as the potentially rich regions
surrounding them, we extended our search area south of the midplane to provide uninterrupted coverage to $b=-40^\circ$.  
Similarly, coverage of the Cepheus molecular cloud was extended north to $b=32^\circ$, and extended southwards to merge with the stripe covering the midplane.  See Fig.~\ref{fig:regions} for details.

\begin{figure*}
\begin{center}
\includegraphics[width=0.75\columnwidth]{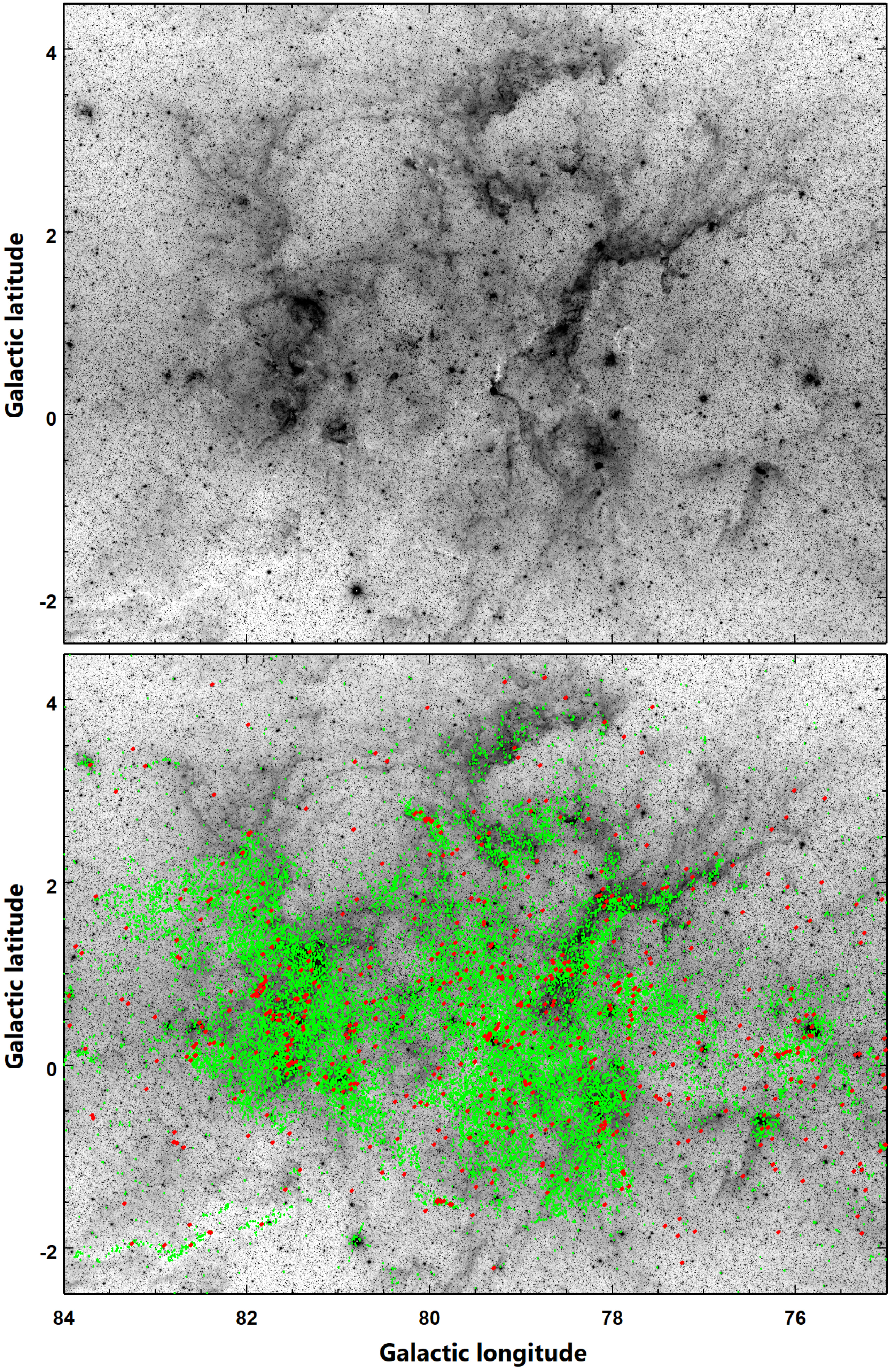}
\end{center}
\caption{An illustration of the density and distribution of \SPHEREx\ Ices targets, showing the Cygnus region as imaged by the Cygnus-X program \citep{2010ApJ...720..679B}.   Coordinates are in degrees of Galactic longitude and latitude.  {\sl Upper panel:} The {\sl WISE} 3.4\,$\mu$m Cygnus X mosaic showing the nebulosity tracing regions of high density in the Milky Way interstellar medium.  
{\sl Lower panel:}  The same, but overlaid with approximately 24500 green dots indicating the positions of SPLICES targets having $H-W2>1.55$,
i.e., the targets thought to lie behind $A_V>10$\,mag of extinction.  Also shown (in red) are the positions of another 600 $H$-undetected targets satisfying the $K_s-W2$ color criterion; these are candidate deeply embedded objects.  The SPLICES targets in this field tend to lie in regions of high nebulosity, although some sources evidently lie distant from the nebulosity.
}
\label{fig:cygnus}
\end{figure*}

\subsection{Target Brightness Criteria} \label{ssec:Thresholds}

All \SPHEREx\ Ices targets are selected using a combination of 2MASS and {\sl WISE} broadband photometry.
By virtue of the red color selection technique, the {\sl WISE W2} band  is likely to contain the most flux of any of 
these filters.  To ensure that SPLICES spectra would have adequate S/N to support reliable interpretations of  
ice absorption features, we imposed a faint-source flux threshold in {\sl W2}.   Using {\sl W2} is arguably the most conservative approach because it is unlikely to unintentionally exclude bright sources in the \SPHEREx\ Bands where ice features of interest lie.  $W1$ in particular is not as useful as an unbiased selection band, because it contains the very broad and deep 3.0\,$\mu$m H$_2$O water absorption feature.

The specific requirement adopted to exclude faint sources is $W2<11.94$\,Vega mag, 
set to match the estimated \SPHEREx\ 20$\sigma$ per-channel 
point-source sensitivity in Band 6 
following expectations during Phase C of mission development \citep{Crill.2020}.  In other words, 
sources fainter than $W2=11.94$\,Vega mag (15.28\,AB mag) were excluded from \acronym.  

To summarize: the SPLICES \version\ selection criteria consist of a $$W2<11.94\,{\rm mag}$$ brightness threshold combined with a color criterion that is either,
$$H-W2>0.324\,{\rm mag},$$ or (if the target lacks a 2MASS PSC $H$ magnitude with a quality flag ``A" or ``B", but was detected in the $K_s$ band with a quality flag of ``A") 
$$K_s-W2>0.55\,{\rm mag},$$ plus a 6\farcs2 isolation criterion imposed to reduce or eliminate spectral contamination.  Finally, only sources lying inside the regions described in Sec.~\ref{sec:regions} are included.  

\begin{figure*}
\begin{center}
\includegraphics[width=1.0\columnwidth]{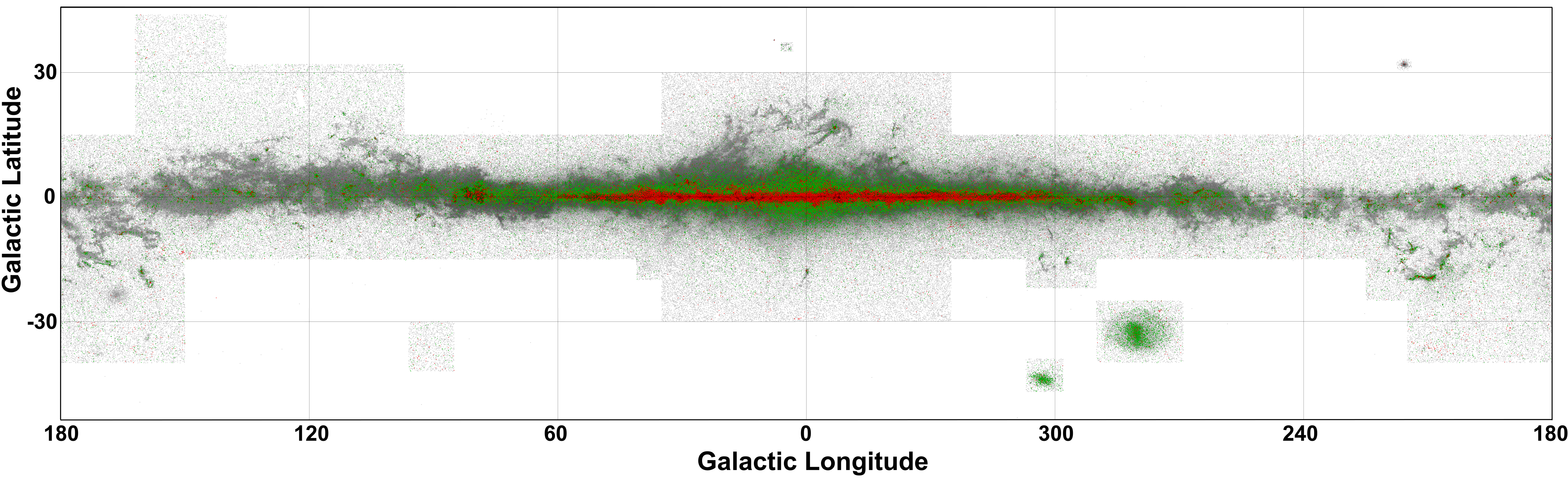}
\end{center}
\caption{The spatial distribution of SPLICES targets in Galactic coordinates.  Despite the cloudlike appearance of this image, it consists of more than eight million discrete dots each indicating the position of a source satisfying the SPLICES isolation, brightness, and color criteria described in Sec.~\ref{sec:selection}.  Red dots indicate the locations of $H$-undetected objects selected on the basis of their red $K_s-W2$ colors.  Green dots indicate targets expected to lie behind $A_V>10$\,mag of dust extinction, based on their $H-W2$ colors.  Unsurprisingly, the distribution traces the structure of known Milky Way molecular clouds, and concentrates toward the central $\pm60^\circ$ of the Milky Way disk.  
 }
\label{fig:spatial_dist}
\end{figure*}

\begin{figure}
\begin{center}
\includegraphics[width=0.5\columnwidth]{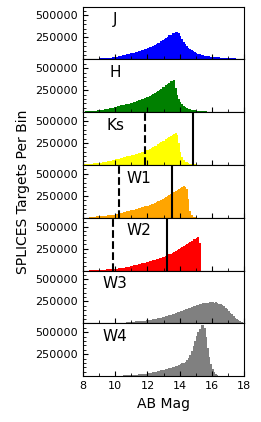}
\end{center}
\caption{The SPLICES target brightness distributions in the 2MASS and {\sl WISE} bands, using 0.1\,mag bins.  The  solid vertical lines indicate estimated SNR=50 detections per \SPHEREx\ spectral spectral resolution element in Bands 3, 5, and 6 (in the panels for $K_s$, $W1$, and $W2$, respectively, \citealt{Crill.2020}).   The dashed lines indicate the magnitudes 
at which the onset of saturation occurs, as defined by the need to apply a nonlinearity correction, given the present
understanding of the detector electronics.  In principle sources
 this bright ought to yield recoverable spectra.  
 Spectra for sources more than $\sim1$\,mag brighter than magnitudes indicated by the dashed
lines, however, are likely unrecoverable at those wavelengths.  }
\label{fig:mag_dist}
\end{figure}

The distribution and abundance of SPLICES targets identified using the color, isolation, and coverage criteria described above are shown within the Cygnus region in particular in Fig.~\ref{fig:cygnus}.
The spatial distributions of SPLICES targets are indicated on a larger scale in Fig.~\ref{fig:spatial_dist}.
The brightness distributions of ices targets are shown in Fig.~\ref{fig:mag_dist}.

\section{Additional Photometry from Cross-Matching} \label{sec:xmatching}

Sources within the survey regions described above that satisfy the target selection criteria defined in Sec.~\ref{sec:selection} were cross-matched to other catalogs in order to fill out their spectral energy distributions.

We performed a proximity search for 3.6, 4.5, 5.8, and 8.0\,$\mu$m counterparts in regions that had been surveyed with {\sl Spitzer}'s InfraRed Array Camera (IRAC; \citealt{2004ApJS..154...10F}).   In the central region of the Milky Way midplane, that 
included a compilation of Galactic Legacy Infrared Mid-Plane Surveys Extraordinaire (GLIMPSE; \citealt{2009PASP..121..213C}).  At greater Galactic longitudes, a combination of GLIMPSE360, Deep GLIMPSE, SMOG, and the Cygnus-X and Vela-Carina surveys were used (\citealt{2020AJ....160...68W}; 
\citealt{2019ApJ...880....9W}; \citealt{2010ApJ...720..679B}; \citealt{2009ApJ...707..510Z}).  It was necessary to combine numerous surveys to cover the entire midplane.  For fields at high Galactic latitude, single catalogs were sufficient.  Targets in the LMC and SMC were respectively matched to the Spitzer Survey of the Large Magellanic Cloud (SAGE; 
\citealt{2006AJ....132.2268M}) and the Spitzer Survey of the Small Magellanic Cloud (SAGE-SMC; 
\citealt{2011AJ....142..102G}).  In all cases the matching was done using a 1\farcs5 search radius and the nearest IRAC source was taken to be the correct match.
The resulting matched photometry is somewhat heterogeneous in terms of both sensitivity and bands covered.  The depths vary significantly from field to field, and only surveys undertaken during {\sl Spitzer}'s cryogenic mission, such as GLIMPSE and SAGE, 
benefit from coverage in the long-wavelength IRAC bands. 

After completing the proximity matching to IRAC-detected sources, the 2MASS counterparts drawn from the 
IRAC-based catalogs were cross-checked against the 2MASS PSC objects used for the initial color-based 
target selection.  In some cases the 2MASS designations differed slightly, but visual inspection revealed 
that they corresponded to the same source.  The 2MASS identifications, which are position-based, 
suggested subarcsecond-level coordinate discrepancies for such targets.  For all these objects we retained 
the 2MASS PSC identifications and cataloged properties originally generated by the AllWISE-2MASS 
proximity matching.

\section{Ad Hoc Catalog Additions}\label{sec:adhoc}

Naturally, not all sources of interest to the \SPHEREx\ Ices investigation lie within the regions outlined 
in Fig.~\ref{fig:regions}, and there are other relevant targets that do not satisfy the broadband selection criteria described in Sec.~\ref{sec:selection}.  To ensure that at least some of these sources are not missed, they have been added to \acronym\ on an ad hoc basis, as described below.

\subsection{Circumstellar Disks}\label{ssec:disks}

In addition to the uniform color selection described above, a literature search was performed to identify well-known protoplanetary disks so as to include this important class of early-stage objects in SPLICES.  A total of 91 disks were identified in a private compilation (K. Pontoppidan, priv. comm.), which drew from, e.g., \citet{2009ApJS..180...84W, 2010ApJ...720..887P, 2013ApJ...770...94B, 2017ApJ...844...99L, 2018ApJ...869L..41A, 2019ApJ...872..158A}, and others.  We combined that list with that curated at {\tt circumstellardisks.org} as of the 2021 August 13 update.  After eliminating sources common to both, and highly extended sources unsuitable for automated forced photometry in the manner planned for the SPHEREx pipeline, we cross-checked the resulting combined list of 360 unique disk sources against SPLICES.  Unlike the sources selected as described in Sec.~\ref{sec:selection} the disks were not screened to eliminate sources with nearby companions, because there are relatively few of them: they tend to be bright, high-value sources.

Of the 360 sources in the combined disk list, 135 were already present in SPLICES as a result of the uniform selection carried out as described in Sec.~\ref{sec:selection}, and 225 were absent.  Many of the 225 `missing' disks are either saturated in the AllWISE catalog, not identified there as a discrete source due to either significant extended emission or to substantial confusion from nearby objects, or are fainter than the {\sl W2} cutoff magnitude.  A significant fraction (86/225) of the missing sources, however,
 lie outside the regions previously searched for obscured sources (Fig. 1).  Of the 274 circumstellar disks within the search regions, the 135 selected according to our color criteria represent a $\sim$49\% selection efficiency for this very heterogeneously assembled sample.  Thus, unknown circumstellar disks which satisfy our color criteria have a roughly one-in-two chance of being included in SPLICES.

All the disk sources added to SPLICES are flagged as such as described in Sec.~\ref{sec:content}.

\subsection{Milky Way Stars in Open Clusters}\label{ssec:cluster_stars}

Open star clusters provide many potential targets of interest even though they may not lie behind significant columns of obscuring dust.  Several offer numerous spatially well-separated and often well-studied members, residing at (virtually identical) known distances, with very similar ages.  We therefore included stars from three open star clusters in \acronym\ as a means of validating the performance of the science data reduction pipeline.  

{\sl Hyades:}  Two teams working independently \citep{2019A&A...621L...2R,2019A&A...621L...3M} recently reported robust {\sl Gaia} DR2-based stellar membership lists for the Hyades.  They also documented tidal tails for this open star cluster which extends over 100\,deg.  For the purposes of the present work we adopt the \citet{2019A&A...621L...3M} selection and incorporate their 238 secure sources into SPLICES.

{\sl Pleiades:} We included Pleiades stars identified using an astrometric membership selection (P.\ Cargile, priv.\ comm.).  They were drawn from a parent sample consisting of all {\sl Gaia} DR3 stars within 3 times the \citet{2018A&A...615A..49C} cluster radius.  The parent sample's parallaxes and proper motions were analyzed with a HDBscan cluster finding algorithm \citep{2017JOSS....2..205M} to subsequently identify the Pleiades as the most significant cluster in the astrometric data. 
The 1506 stars with derived HDBscan membership probabilities $> 0.0$ were taken as Pleiades members and included in SPLICES.

{\sl M67:} \citet{2021MNRAS.502.2582A} analyzed {\sl Gaia} DR2 proper motions and parallax measurements 
with a machine learning algorithm in the vicinity of M67 in order to probabilistically assign likelihoods of 
cluster membership to stars brighter than $G\sim20$\,AB mag within roughly 100\arcmin\ of the cluster center.   \citet{2021MNRAS.502.2582A} estimate a total level of contamination under 2\% for their M67 sample.  
Here we include all of the roughly 1200 members of M67 identified by \citet{2021MNRAS.502.2582A}.

For all three clusters considered here (the Hyades, the Pleiades, and M67), we substitute {\sl Gaia} DR3 astrometry and photometry for the older {\sl Gaia} data on which the original selections were based.

\subsection{Other Milky Way Stars}\label{ssec:mwstars}

Some additional well-characterized stars have been included in order to provide an independent means of validating spectral modeling efforts.

{\sl Kepler:} The {\sl Kepler} mission spent years observing a dedicated field in Cygnus; one among many outcomes of that mission was the construction of a sample of Sun-like stars \citep{2017ApJ...835..172L}.  This sample, the {\sl Kepler} Asteroseismic LEGACY Sample, consists of 66 stars of roughly Solar mass, with very well characterized effective temperatures and surface gravities.  

{\sl Gaia Benchmark Stars:}
We examined the {\sl Gaia} FGK Benchmark star sample \citep{2015A&A...582A..49H}, a set of 34 stars covering a range of effective temperatures and having an abundance of spectroscopy in the literature, hence, very accurate stellar parameters.  They are also bright, making them suitable for accurate distance determinations with {\sl Gaia}.  Unfortunately for  our purposes, most of the {\sl Gaia} FGK stars are so bright that they are expected to saturate the \SPHEREx\ arrays.  We therefore included only five of the \citet{2015A&A...582A..49H} stars in SPLICES: HD84937, HD102200, HD106038, HD140283, and HD298986.

{\sl Spitzer/IRAC Calibrator Stars:}
Several of the CALSPEC A stars were used to calibrate {\sl Spitzer}'s Infrared Array Camera.  To those we add the K stars from  \citet{2005PASP..117..978R}.

\section{Catalog Validation}\label{sec:validation}

In this Section some specific examples are provided to illustrate the effectiveness and limitations of the SPLICES target selection criteria.

\citet{2021ApJ...916...75O} describe two bright objects showing strong ice absorption features in {\sl AKARI} mid-infrared spectra.  Both objects were covered in the {\sl AKARI}/Infrared Camera survey of the Galactic plane, and thus fall within the search region defined for the \SPHEREx\ Ices Investigation.  
Both sources in \citet{2021ApJ...916...75O} also present red colors that satisfy the selection criteria in 
Sec.~\ref{sec:selection}.   They are both present in the present version of SPLICES.  
We note, however, that based on current best estimates of instrument performance \citep{Crill.2020}, both these 
very bright targets 
are likely to strongly saturate  \SPHEREx\ spectral channels at wavelengths longward of roughly 2\,$\mu$m.

\subsection{Ad Hoc YSO Search}\label{ssec:adhocyso}

To more systematically investigate the effectiveness of the adopted color criteria (Sec.~\ref{ssec:color_criteria}) for identifying targets of interest, an extensive survey of the refereed YSO literature in particular was undertaken.  
All relevant catalogs in VizieR were identified by a search using the YSOs keyword.  
This search identified a total of 42 VizieR catalogs containing at least some YSOs not 
already present in SPLICES (Table~\ref{tab:vizier_ysos}).  All targets in the 42 catalogs 
were matched to AllWISE counterparts using a 1\farcs5 search radius.  
In total, the 42 studies contain 217398 YSOs satisfying the SPLICES brightness criterion ($W2<11.94$\,mag).   
Removing non-YSO targets (i.e., objects classified as galaxies, stars, contamination, No Class, uncertain, field star, and so on), reduces the total number of potential targets to 51595.

\begin{deluxetable*}{rlrrr}
\tablecaption{Ad Hoc Additions of YSOs from Recent Publications to SPLICES\label{tab:vizier_ysos}}
\renewcommand{\arraystretch}{0.9}
\tablewidth{-2pt}
\tablehead{
\colhead{Ref.}  & \colhead{Reference}  & \colhead{Already In} & \colhead{Added to} & \colhead{Detectable in}\\
\colhead{Code}  & & \colhead{SPLICES}  & \colhead{SPLICES} & \colhead{{\sl WISE W2}} \\
\colhead{(1)}             & \colhead{(2)}  & \colhead{(3)} & \colhead{(4)} & \colhead{(5)} 
                 }
\startdata
 1 & \cite{Kuhn.2021} & 14139 &  1284 & 18112 \\ 
 2 & \cite{2020AJ....160...68W} & 11208 &   651 & 12536 \\ 
 3 & \cite{Dunham.2015} &  1682 &    56 &  1864 \\ 
 4 & \cite{Megeath.2012} &  1484 &    50 &  1570 \\ 
 5 & \cite{2010ApJ...720..679B,DVN/1QMXB8_2021} &   998 &    89 &  1252 \\ 
 6 & \cite{Rebull.2011} &   548 &    48 &   619 \\ 
 7 & \cite{Cottle.2018} &   539 &     7 &   548 \\ 
 8 & \cite{Saral.2017} &   290 &    40 &   431 \\ 
 9 & \cite{Allen.2012} &   383 &    27 &   428 \\ 
10 & \cite{Gutermuth.2009} &   372 &    18 &   406 \\ 
11 & \cite{Esplin.2019} &   318 &    19 &   349 \\ 
12 & \cite{Retes-Romero.2017} &   244 &     4 &   280 \\ 
13 & \cite{Fang.2013} &   207 &    57 &   266 \\ 
14 & \cite{Venuti.2018} &   178 &    57 &   239 \\ 
15 & \cite{Mintz.2021} &   214 &     3 &   225 \\ 
16 & \cite{Azimlu.2015} &   102 &    52 &   179 \\ 
17 & \cite{Chavarria.2014} &   109 &    25 &   140 \\ 
18 & \cite{Grossschedl.2019} &   118 &    17 &   137 \\ 
19 & \cite{Billot.2010} &   125 &     2 &   132 \\ 
20 & \cite{Mallick.2013} &   111 &     5 &   127 \\ 
21 & \cite{Fischer.2016} &   119 &     4 &   125 \\ 
22 & \cite{Riviere-Marichalar.2016} &    45 &    34 &   110 \\ 
23 & \cite{Rapson.2014} &   100 &     5 &   109 \\ 
24 & \cite{Koenig.2015} &   100 &     1 &   101 \\ 
25 & \cite{Rivera-Ingraham.2011} &    67 &    24 &    97 \\ 
26 & \cite{Gama.2016} &    79 &     5 &    95 \\ 
27 & \cite{Willis.2013} &    77 &     7 &    94 \\ 
28 & \cite{Cambresy.2013} &    73 &    13 &    87 \\ 
29 & \cite{Kim.2015} &    67 &     2 &    80 \\ 
30 & \cite{Flaherty.2013} &    75 &     2 &    77 \\ 
31 & \cite{Kun.2016} &    66 &     6 &    72 \\ 
32 & \cite{Connelley.2008} &    47 &     4 &    57 \\ 
33 & \cite{Rebull.2020} &    21 &    29 &    50 \\ 
34 & \cite{Koenig.2008} &    27 &    18 &    48 \\ 
35 & \cite{Saral.2015} &    32 &     1 &    43 \\ 
36 & \cite{Dent.2013} &    16 &    27 &    43 \\ 
37 & \cite{Ragan.2009} &    30 &     2 &    42 \\ 
38 & \cite{Kounkel.2014} &    26 &    16 &    42 \\ 
39 & \cite{Wolk.2015} &    25 &     9 &    34 \\ 
40 & \cite{Rebull.2015} &    26 &     1 &    28 \\ 
41 & \cite{Morales-Calderon.2009} &    21 &     2 &    26 \\ 
42 & \cite{Carlson.2012} &     4 &     4 &     8 \\ 
& Total &  34512 & 2727 & 41308 \\
\enddata
\tablecomments{YSOs added to SPLICES from the literature search described in Sec.~\ref{sec:validation}.   Col.\ 1: Rank of catalog.  Col.\ 2: VizierR citation.  Col.\ 3: targets already present in SPLICES.  Col.\ 4: targets not present in SPLICES nor in previously examined studies, and which were added to SPLICES.  Col.\ 5: targets brighter than $W2=11.94$\,mag 
according to AllWISE, and which satisfy the SPLICES isolation criterion.}
\end{deluxetable*}

We ranked the
catalogs by size, and then proceeded in rank order to eliminate duplicate targets based on the AllWISE ID, reducing the total number of YSOs to 34512 unique objects.  After removing objects by applying the 6\farcs2 isolation criterion, we identified 2727 unique objects that were not already present in SPLICES.  Table~\ref{tab:vizier_ysos} shows how many sources came from the individual investigations.  
The YSOs compiled in this way have been added to SPLICES and
are flagged as ad hoc additions.   Where the authors provide YSO classes explicitly, we respect their classifications, for the existing sources as well as those added to SPLICES.   As a result, 
6151 protostars (Class 0/I),
4168 flat-spectrum sources, 
24458 disk sources, and 
6557 evolved sources (Class III) are marked as such in SPLICES.  
However, when no classifications were provided we classify the targets ourselves
 on the basis of the measured infrared spectral indices $\alpha$ following \cite{1994ApJ...434..614G}: $\alpha > 0.3$ as Class I, $0.3 > \alpha \geq -0.3$ as ``flat spectrum'', $-0.3 > \alpha \geq -1.6$ as Class II, and $\alpha < -1.6$ as Class III.   

In summary, a thorough literature search added $\sim3\times10^3$ new science targets to a list that had previously contained order $10^7$ targets, suggesting that relatively few isolated YSOs have been missed in compiling SPLICES. It is therefore anticipated that \SPHEREx\ observations will reliably sample the ranges of phenomena exhibited by bright YSOs.

\subsection{IRTF Spectroscopy of Bright SPLICES Analogues}\label{ssec:IRTF}

Contamination of SPLICES by intrinsically red targets is a potential problem.  Specifically, active galactic nuclei (AGN), well known for their red IR colors (e.g., \citealt{2005ApJ...631..163S, 2007AJ....133..186L}) or evolved stars \citep{2022arXiv220607668L} can both in principle satisfy all the selection criteria adopted here to build a list of ostensibly 
obscured sources.  We therefore chose to test our selection scheme with ground-based spectroscopy, so as to
better understand the character and content of SPLICES before \SPHEREx\ launches.

In the first semester of 2022 we were awarded IRTF time for a pilot study to obtain 2.8--4.2\,$\mu$m spectra with SpeX 
(a $R\sim$2500 mid-IR spectrometer; \citealt{2003PASP..115..362R})  
 of about two dozen bright targets on the partial nights 
of 2022 June 11, 20, and 21 (program 2022A065, PI Ashby).  
All objects targeted for program 2022A065 were very bright, to an extreme that would 
saturate the SPHEREx long-wavelength arrays in flight; they belong to the bright tail 
of the $W2$ distribution visible in Fig.~\ref{fig:mag_dist}.
This was of course necessary for ground-based mid-IR spectroscopy.  
 
On the three working IRTF nights a total of 22 sources were observed, spanning ranges of $H-W2$ color 
(our proxy for $A_V$) and Galactic latitude (to test the degree to which contamination by evolved 
stars and AGN depends on latitude).  
Observations of targets were interleaved with telluric standard HD\,192538 (an A0V star),
 and with pre-programmed calibration sequences.  
The spectra were reduced in the standard way using Spextool (Cushing, Vacca, \& Rayner 2014).

\begin{figure}
\begin{center}
\includegraphics[width=0.5\columnwidth]{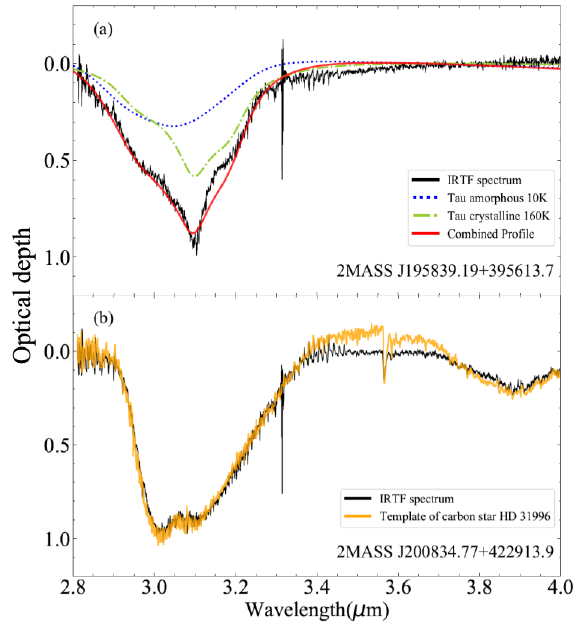}
\end{center}
\caption{IRTF/SpeX spectra of two extremely bright SPLICES targets acquired during our 2022A 
pilot survey to validate the target selection.  {\sl Top:} 2MASS J195839.19+395613.7 (in black), exhibiting a strong water ice feature at 3.0\,$\mu$m.  {\sl Bottom:} 2MASS J200834.77+422913.9 (in black), which is best fit by an IRTF spectral template from carbon star HD\,31996 (yellow).
  }
\label{fig:IRTF_spectra}
\end{figure}

Four sources of the 22 observed presented strong absorption features at 3\,$\mu$m.  Of these, the features for three were not well-fitted by any combination of amorphous and crystalline ice that we attempted.  Indeed, their
3\,$\mu$m absorption features were found to be consistent with those seen in carbon stars rather than water ice; one of them is
compared to a carbon star template in the lower panel of Fig.~\ref{fig:IRTF_spectra}.  The fourth source, 2MASS 
J195839.19+395613.7 (upper panel of Fig.~\ref{fig:IRTF_spectra}), had a 3.0\,$\mu$m absorption feature consistent with a dust-reddened combination of
amorphous and crystalline water ice.  
None of the ten observed targets with estimated $A_V\le6$ showed any evidence for absorption features.
While it is premature to draw firm conclusions about the SPLICES sample as a whole from such limited results,
there are hints in these initial observations that, first, carbon stars may contaminate the selection (at least for bright magnitudes where carbon stars are well known) to some degree, and second, that ice features may tend to preferentially occur toward targets with higher proxies for $A_V$.  We have submitted a follow up IRTF/SpeX proposal to characterize the degree of carbon star contamination specifically in SPLICES in the coming semester.

\subsection{NEOWise Variability Analysis}\label{sec:variability}

Given its orbit and Sun avoidance requirements, {\sl SPHEREx} will cover most sky positions 
at approximately six-month intervals.  Thus, the intrinsic variability of background targets on these or indeed any 
timescale will potentially impact the usefulness of spectra returned by the mission; variability on short timescales in
particular can mimic spectral features.  Fortunately, there exists a public database of reliable, 
long-baseline mid-IR photometry generated by the {\sl NEOWISE} mission that has been used to identify many variable sources in SPLICES, which we describe here.

After completing its initial mission, {\sl WISE} \citep{2010AJ....140.1868W} was reactivated for four months
\citep{2011Mainzer} during which it acquired {\sl W1} and {\sl W2} photometry.  In 2013 September the spacecraft 
was again reactivated as {\sl NEOWISE-R}, the NEOWISE-reactivation Post-Cryogenic Mission \citep{2014Mainzer}, again using only the {\sl W1} and {\sl W2} bands, as indeed it continues operating 
to the present.  To identify variable sources in SPLICES and thereby prevent spurious measurements 
of spectral features in SPLICES targets, we have analyzed the resulting {\sl NEOWISE}
archive of multi-epoch {\sl W1}+{\sl W2} photometry up to and including the observations 
of mid-December, 2021.  The {\sl WISE} and {\sl NEOWISE} archives were searched for all photometry 
within 3\arcsec\ from the coordinates of every SPLICES target. 
Averaging of the multi-epoch data was done with the same method as \citet{2021ApJ...920..132P}.

Variable SPLICES targets were classified following the methods of \citet{2021ApJ...920..132P}, 
who characterized long-term secular and stochastic variability in $\sim$7000 YSOs 
in nearby star-forming regions, using only the {\sl W2} band.
 Like \citet{2021ApJ...920..132P} we identified objects as variable if they satisfied the condition 
 $\Delta W1/\sigma_{W1}\geq3$ or $\Delta W2/\sigma_{W2}\geq3$, where
 $\sigma_{W1}$ and $\sigma_{W2}$ indicate the variance measured in their respective bands after rejection of artifacts.  Lomb-Scargle \citep{1976Lomb,1989Scargle} periodogram and linear fits were used to search for secular trends in the light curves of variable objects, permitting classification of the variable sources based on their behavior.  YSOs well-fitted by a linear model are classified as {\it linear} variables.  YSOs not following linear trends, but well-fitted by a sinusoidal light curve are classified as {\it periodic}, if they have periods of $P<P_{lim}$, or {\it curved} if their periods are longer than $P_{lim}$, where $P_{lim}$ is defined as half of the time covered by the observations.  In \citet{2021ApJ...920..132P}, $P_{lim}=1200$\,d.  Here $P_{lim}=1400$\,d, i.e., half of the available temporal baseline.   No variability with a period shorter than 200\,d was considered.  
 YSOs that do not fall into the secular variability classes are defined as showing stochastic variability. 
 The latter are further divided into {\it burst}, {\it drop} and {\it irregular} classes, based on the shape of the light curves. 
 For further details on the classification criteria see Sections 3 and 4 of \citet{2021ApJ...920..132P}.

The variability classification of SPLICEs targets was performed using the {\sl WISE} {\sl W1} and {\sl W2} bands independently.  For each source the catalogue contains the average magnitude, amplitude and variability class in each filter. For objects that are classified as periodic or curved we also provide the best period estimated by the Lomb-Scargle periodogram.  The periods provided in this analysis should be taken as approximate only.

Including the {\sl WISE} and {\sl NEOWISE} surveys, fully $>97\%$ of the SPLICES objects have 16 or more epochs of photometry in both {\sl W1} and {\sl W2} filters.  From these measurements, about 7-9\% of SPLICES meet the variability criteria.  Inspection of a sub-sample of light curves shows that among variable sources different types of systems are present including protostars, Active Galactic Nuclei and Asymptotic Giant Branch Stars.  The variability analysis will be described in detail by C.\,Contreras et al.\ in prep.

\begin{figure*}
{\includegraphics[width=0.5\columnwidth]{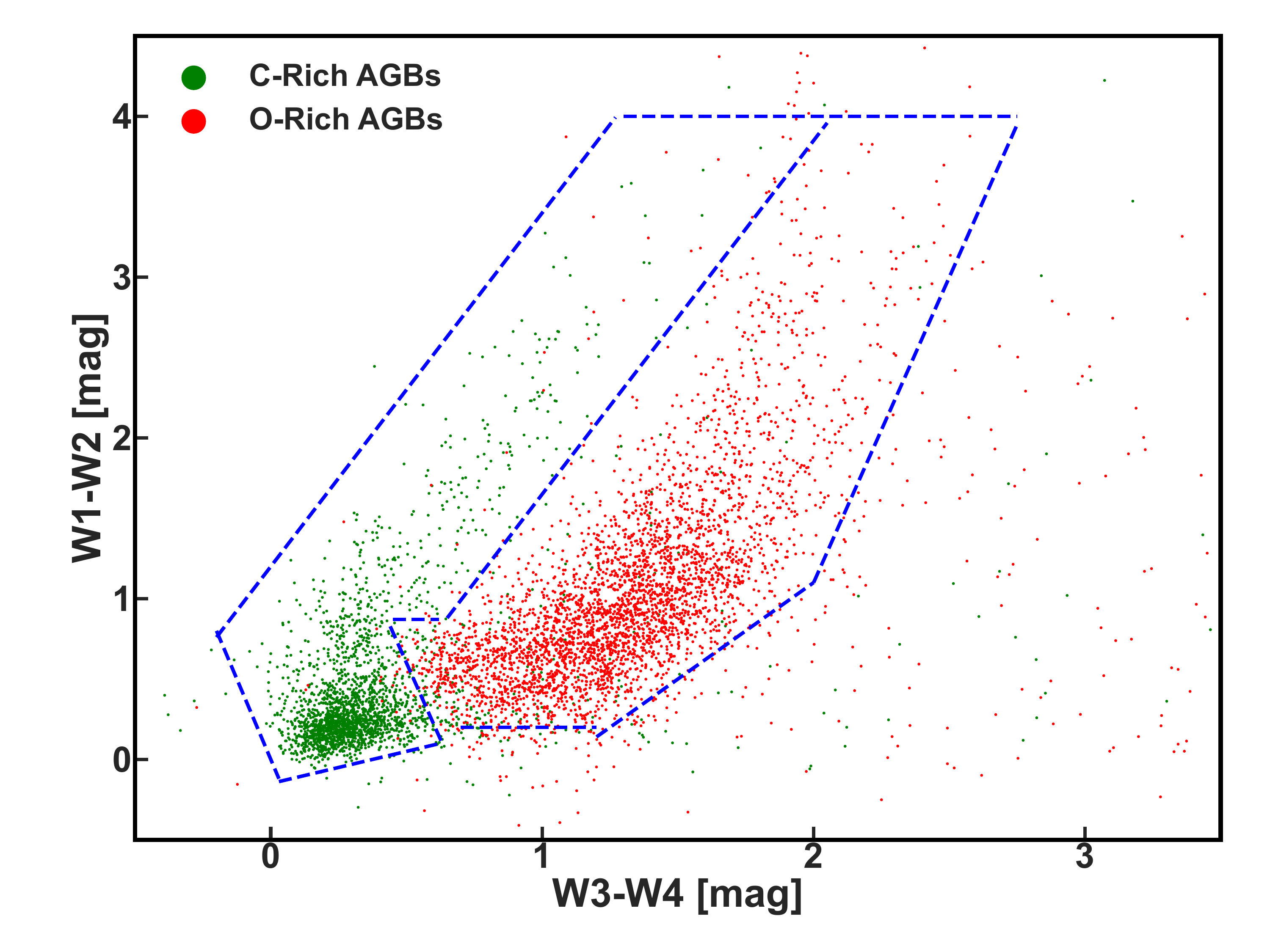}}
{\includegraphics[width=0.5\columnwidth]{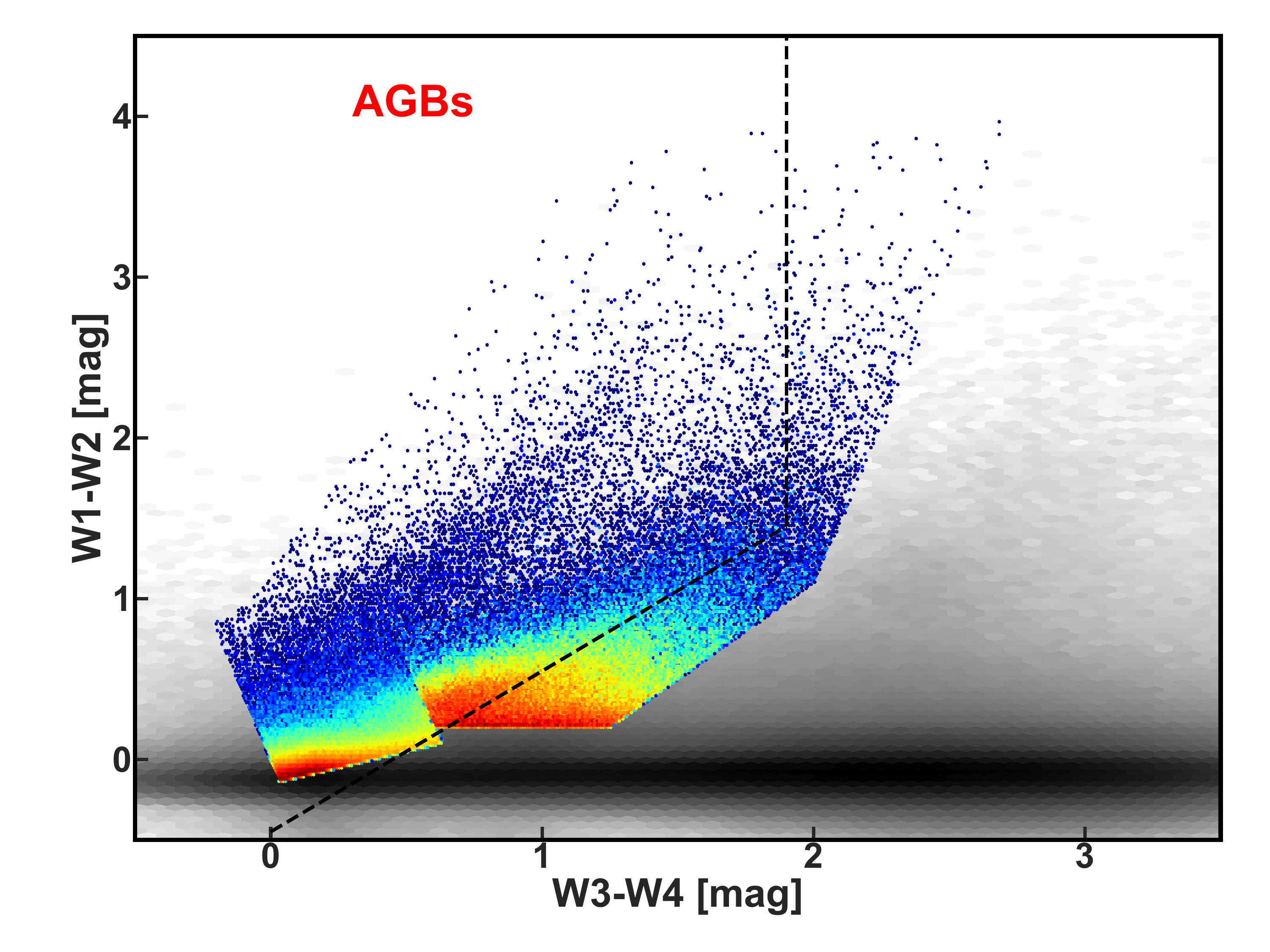}}\\
{\includegraphics[width=0.5\columnwidth]{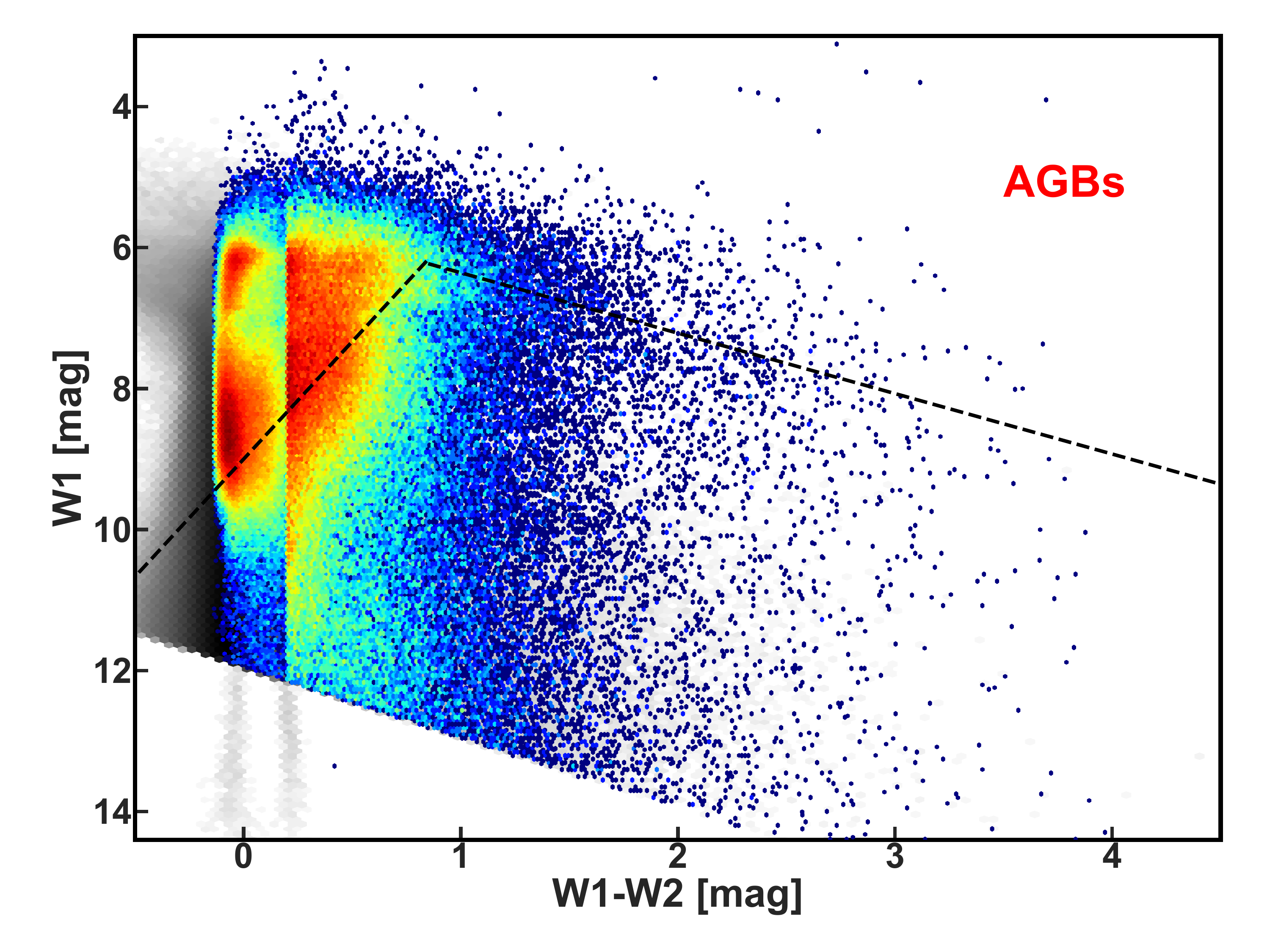}}
{\includegraphics[width=0.5\columnwidth]{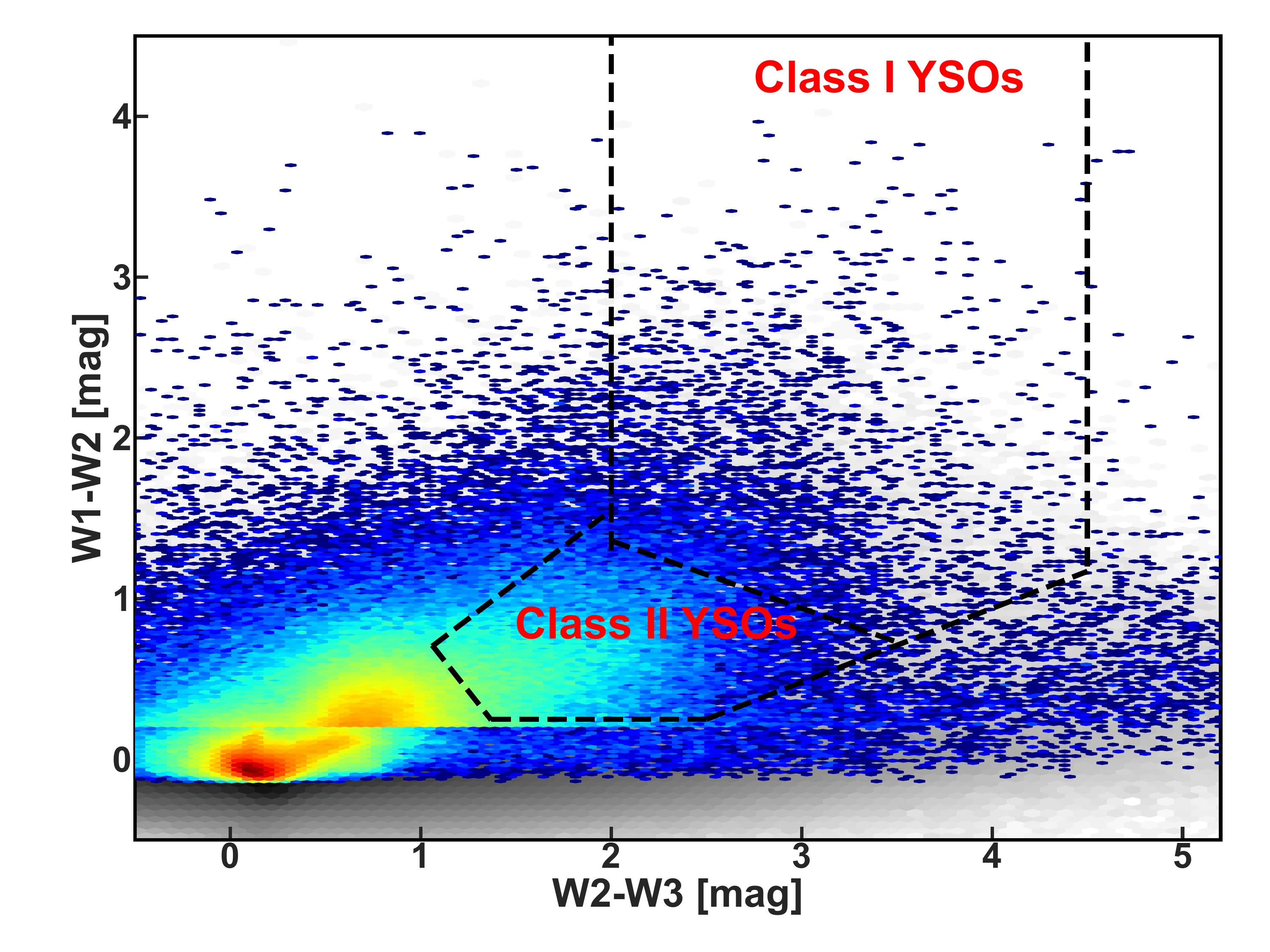}}
  \caption{Source classification diagrams based on {\sl WISE} broadband colors.  {\sl Upper left:} $W1-W2$ vs $W3-W4$ color-color diagram of known carbon- and oxygen-rich Galactic AGB stars (green and red circles, respectively) from the catalogue of \citet{2021ApJS..256...43S}.  The regions defined to contain most of carbon- and oxygen-rich AGBs are marked by the dashed blue lines. {\sl Upper right:}  $W1-W2$ vs $W3-W4$ color-color diagram of SPLICEs targets (grey) and those classified as AGB stars (color).  Dashed black lines indicate the AGN-dominated regions following \citet{2014Koenig}.  {\sl Lower left:} $W1$ vs $W1-W2$ color-magnitude diagram. Symbols and lines identical to those in the previous panel. {\sl Lower right:} $W1-W2$ vs $W2-W3$ color-color diagram of AGB candidates from SPLICEs.  The regions used by \citet{2014Koenig} to select Class I and Class II YSOs are indicated.}
    \label{fig:AGB1}
\end{figure*}

\subsubsection{Asymptotic Giant Branch Stars}

To assess the potential degree of contamination from carbon-rich evolved stars, we used the $W1-W2$ vs $W3-W4$ color distribution of known Galactic carbon- and oxygen-rich Asymptotic Giant Branch (AGB) stars taken from the catalogue of \citet{2021ApJS..256...43S} because most carbon-rich stars are AGBs.  From visual inspection of 
how the AGB stars distribute in this color-color space, we defined regions that most likely contain C- and O-rich AGBs (Fig.~\ref{fig:AGB1}).  Corresponding color cuts applied to the SPLICES targets indicate that 
1.5$\%$ and 2.4$\%$ of them can be classified as possible oxygen-rich and carbon-rich AGB stars, respectively (top-right panel of Fig.~\ref{fig:AGB1}).

As a check we compared to \citet{2014Koenig} (hereafter KL14), who use {\sl WISE} colors to classify YSOs.  To avoid contamination from evolved stars, KL14 define several regions in color-color and color-magnitude diagrams where AGBs are likely to locate (see the top-right and bottom-left panels of Fig.~\ref{fig:AGB1}).
We find that 86$\%$ of SPLICEs targets classified as AGBs on the basis of their colors in these diagrams would also be classified as AGBs following the KL14 criteria.  If we only consider those objects classified as carbon-rich AGBs, we find that 99.3$\%$ of them would also be classified as AGBs according to KL14.
By contrast, only 4$\%$ of SPLICEs targets classified as AGBs would also be classified as YSOs by the KL14 criteria.  Furthermore, the overlap occurs only for oxygen-rich AGBs because no SPLICES objects classified as carbon-rich AGBs meet the KL14 criteria.

AGBs can also be identified due to their nearly-sinusoidal variability \citep[e.g.,][]{2021Lee}.  Therefore, we used the additional information obtained from the variability analysis to classify objects as periodic AGB candidates if they are classified as periodic in both the {\sl W1} and {\sl W2} bands.  We find that $\sim8\%$ of candidate AGBs selected from the upper left panel of Fig.~\ref{fig:AGB1} are classified as periodic.  These objects are more likely to be true AGB stars. 

Following the above analysis, we provide one additional column in SPLICES that flags sources as candidate AGB stars. The information is provided as {\it CAGB} or {\it OAGB} if they fall in the C-rich or O-rich region respectively. If objects are additionally classified as periodic then they are classified as either {\it CAGBp} or {\it OAGBp}.  

\subsection{Probabilistic Target Classification Using Machine Learning}\label{ssec:ML}

In a companion paper, K.\ Lakshmipathaiah et al.\ in prep.\ (submitted, hereafter KL23) describe  
supervised machine learning algorithms used to classify the SPLICES targets based on
photometry in the five combined 2MASS and AllWISE bands. 
Briefly, KL23 used an ensemble classifier that combined 
Random Forest Classifier, Fully Connected, 
and Convolutional Neural Network techniques to classify the SPLICES targets into
four broad classes:  YSO, AGB star, Active Galactic Nucleus (AGN), and main-sequence (MS) star.  
Sources were assigned to the class with the highest associated probability.  

KL23 also computed the likelihood of the source belonging to its assigned class.  
If the source was identified as a candidate YSO, it was sub-classified
into Class I, Flat spectrum, Class II, and Class III YSO based on its infrared spectral index.
Similarly, using machine learning algorithms, candidate AGB stars were also sub-classified
as carbon-rich (C AGB) and oxygen-rich (O AGB) stars.  The probability associated
with the sub-classification of YSOs and AGB stars was also computed.  Among all
the sources analyzed, 633,814 sources were classified with probability
exceeding 90\%, amongst which YSOs constitute ~56\%, AGB stars ~40\%, and the
remaining are a mixture of (reddened) MS stars and AGNs.  Upon acceptance, 
Column 43 will list the KL23 classifications, while 
Columns 44 and 45 will provide the associated likelihoods.  Details of
the methods adopted and the classifications can be found in KL23, which presents
classifications (and the likelihood thereof) for all our targets classified with
a confidence exceeding 80\%.

\section{Catalog Format}\label{sec:content}

The \SPHEREx\ catalog is available from the Infrared Processing and Analysis Center (\dataset[doi:10.26131/IRSA554]{https://doi.org/<10.26131/IRSA554>}).  It contains approximately \numsources\ candidate sources in total.  
It lists the following information for each source, defined below in column order.  Many of the column descriptions are taken verbatim from the 2MASS, AllWISE, and {\sl Gaia} DR3 documentation:

{\bf Column 1. {\tt designation}}: A 19-character identification string unique to each source, drawn from the AllWISE {\tt designation} field.  

{\bf Columns 2, 3. {\tt ra, dec}}: Right Ascension and Declination in decimal degrees (J2000).   These are best available coordinates.  If {\sl Gaia} identifications have been made, these coordinate are taken from the {\sl Gaia} DR3 catalog.  Otherwise they are drawn from the 2MASS PSC.  They are the coordinates to which all other catalog entries were matched.  

{\bf Columns 4, 5. {\tt GAL\_LONG, GAL\_LAT}}: Galactic longitude and latitude in decimal degrees, computed from the  coordinates in columns 2 and 3.

{\bf Column 6. {\tt tmassdesignation}}: Sexagesimal, equatorial position-based 17-character source name taken directly
from the 2MASS PSC.  The ``2MASS J" prefix is not included.

{\bf Columns 7, 8. {\tt tmassra, tmassdec}}: Right Ascension and Declination in decimal degrees (J2000), drawn from the 2MASS PSC.

{\bf Columns 9, 10. {\tt J, Jerr}}:  The Vega-relative $J$-band magnitude and uncertainty therein, taken from the 2MASS PSC.  The full description is copied here:  Default $J$-band magnitude, or if the source is not detected in the $J$-band, the 95\% confidence upper limit derived from a 4$^{\prime\prime}$ radius aperture measurement taken at the position of the source on the Atlas Image. The origin of the default magnitude is given by the first character of the rd\_flg value (rd\_flg[1]). This column is null if the source is nominally detected in the $J$-band, but no useful brightness estimate could be made (rd\_flg[1]=9).  The corresponding PSC description for $J$err reads as follows:  Combined, or total photometric uncertainty for the default $J$-band magnitude.   In other words, $J$err accounts for measurement and systematic uncertainty.

{\bf  Columns 11, 12, 13, 14. {\tt  H, Herr, K, Kerr}}:  As above for $J$ and $J$err, but these are instead the 2MASS PSC entries corresponding to the $H$ and $K_s$ bands.

{\bf Column 15. {\tt ph\_qual}}:  The 2MASS PSC photometric quality flag, a three-character string.  The first, second, and third characters correspond respectively to the $J$, $H$,  and $K_s$ bands.   The characters can take on several values, but often-appearing characters are A, B, and C, which respectively indicate signal to noise ratio detections of at least 10, 7, and 5 in the corresponding band.  

{\bf Column 16. {\tt rd\_flg}}:  The 2MASS PSC read flag, again a three-character string, which encodes the origin of the default uncertainties and magnitudes in each of the 2MASS bands.  The first, second, and third characters correspond respectively to the $J$, $H$,  and $K_s$ bands.  When the characters are set equal to ``0," that indicates no detection in the corresponding band.  Entries of  ``1," ``2," or ``3" indicate the best photometric measurements.  

{\bf Column 17. {\tt prox:}}  The 2MASS PSC proximity parameter, in decimal arcseconds, to one digit of precision.  This parameter is defined as the distance between this source and its nearest neighbor in the PSC.  Note the PSC documentation claims that the effective resolution of the 2MASS system is approximately 5$^{\prime\prime}$.  Any source with {\tt prox} $<6^{\prime\prime}$ is potentially confused with a nearby object, and should be used with caution.

More information for columns 6-17 is available in the 2MASS PSC documentation.

{\bf Columns 18, 19. {\tt W1, W1err}}: The {\sl WISE} band 1 photometry.  {\tt W1} and {\tt W1err} 
are respectively the Vega-relative 3.4\,$\mu$m magnitude and uncertainty therein from the 
AllWISE band-merged catalog.  

{\bf Columns 20, 21, 22, 23, 24, 25. {\tt W2, W2err, W3, W3err, W4, W4err}}: As above, but for {\sl WISE W2, W3}, and {\sl W4}, respectively.

{\bf Column 26. {\tt cc\_flags}}:  A data quality flag passed through from AllWISE.  {\tt cc\_flags} is a 
contamination and confusion flag.   It is a four-character string, one character per band, in wavelength order.
The single-character code is intended to indicate the following conditions:\\
{\bf D,d}: Diffraction spike.\\
{\bf P,p}: Persistence.\\ 
{\bf H,h}: Halo.\\
{\bf O,o}: (Letter ``o") Optical ghost.\\
{\bf 0}: (Number zero) Source unaffected by known artifacts.\\

{\bf Column 27. {\tt ext\_flg}}: AllWISE extended source flag. This is an integer flag, the value of which indicates whether or not the morphology of a source is consistent with the WISE point spread function in any band, or whether the source is associated with or superimposed on a previously known extended object from the 2MASS Extended Source Catalog (XSC). A value of zero indicates the source shape is consistent with it being a point source.

{\bf Column 28. {\tt var\_flg}}: AllWISE photometric variability flag.  Like {\tt cc\_flags}, this is a four-character string, one character per band, in wavelength order, indicating likelihood of this source being variable.  The likelihood increases as the character increments from ``0" to ``9."  {\tt var\_flg} is based on flux measurements made in
individual exposures, and can therefore indicate either short- or long-term variability, or both.

Columns 18-28 are drawn without modification from AllWISE.

{\bf Column 29. {\tt IRAC\_designation}}: Sexagesimal, equatorial position-based 27-character source name taken directly from the IRAC catalog for which the source was matched using a 1\farcs5 matching radius as described in the main text.

{\bf Columns 30, 31. {\tt ra\_IRAC, dec\_IRAC}}: Right Ascension and Declination in decimal degrees (J2000), drawn from the IRAC catalog containing this source.  

{\bf Columns 32, 33. {\tt mag3\_6, d3\_6m}}:  3.6\,$\mu$m Vega-relative magnitude and uncertainty therein, drawn from the IRAC catalog containing this source.  

  {\bf Columns 34, 35, 36, 37, 38, 39. {\tt mag4\_5, d4\_5m, mag5\_8, d5\_8m, mag8\_0, d8\_0m}}: As {\tt mag3\_6, d3\_6m}, but respectively for the IRAC 4.5, 5.8, and 8.0\,$\mu$m bands.
  
{\bf Column 40. {\tt SEL\_TYPE}}:  A character string indicating the selection under which this source was chosen for inclusion in SPLICES.  Possibilities for {\tt SEL\_TYPE} entries are:
\begin{itemize}
\item MAIN: this object was selected following the main $H-W2$ color criterion, plus the search-area, isolation, and brightness criteria described in Sec.~\ref{sec:selection}.
\item EMBED: similar to MAIN, but satisfying instead the $K_s-W2$ color criterion.
\item ADHOCYSO: a member of the ad hoc YSO selection described in Sec.~\ref{ssec:adhocyso}.
\item DISK: drawn from the circumstellar disk compilation described in Sec.~\ref{ssec:disks}.
\item HYADES, PLEIADES, or M67: a member of one of these three open star clusters, as described in Sec.~\ref{ssec:cluster_stars}.
\end{itemize}

  {\bf Column 41. {\tt CLASS}}: A source classification drawn from the literature associated with the target (see column YSO\_PUB for the reference). Not all targets are classified.  The following scheme has been adopted to unify the variety of nomenclatures in the literature, as described in Sec.~\ref{ssec:adhocyso}: 
\begin{itemize}
\item YSO-P: a protostar, variously classified in the original citations as Class 0/I, ClassI, CI, 1, I, P, ICI, 0/I, and 0.
\item YSO-F: flat-spectrum sources: FS or flat.  
\item YSO-D: disk sources,  i.e.,  ClassII,  classII,  II,  2, CII, D, AD, or TD.
\item YSO-E: evolved sources, i.e., ClassIII, 3, III,  or III/F.
\end{itemize}

  {\bf Column 42. {\tt YSO\_PUB}}: An integer indicating the reference used to classify the target.  Numbers correspond to the references listed in Table~\ref{tab:vizier_ysos}.

  {\bf Column 43. {\tt STYPE\_ML}}: A character string up to 8 characters long indicating the likely nature of the background source, if one has been identified.  Sources are required to be identified with at least 80\% confidence by the machine learning algorithm described in KL23.  The classifications are as follows:  

\begin{itemize}
\item YSO-E, YSO-D, YSO-F, YSO-P: the object is classified as a YSO by the stage-1 ML classifier.  It is also more finely classified into YSO subclasses based on spectral index by the stage-2 ML classifier, following the same scheme as described in Sec.~\ref{ssec:adhocyso}.
\item AGN, MS: the object is classified as an AGN or main sequence star by the stage-1 ML classifier.
\item OAGB, CAGB: the object is classified as an AGB star by the stage-1 ML classifier.  It is also more finely classified as either an oxygen or a carbon star by the stage-2 ML classifier.
\end{itemize}
 
{\bf Column 44. {\tt STYPE\_ML\_PROB1}}:  The percentage likelihood associated specifically with {\tt STYPE\_ML}, ranging from 0 to 100.  The default value is 0, meaning no machine learning-based classification  has been made at the 80\% confidence level or higher.  For YSOs and AGB stars, this column describes the machine learning classifications only broadly, i.e., not considering sub-classifications into, e.g., Carbon or Oxygen AGB stars, or particular types of YSOs.
    
{\bf Column 45. {\tt STYPE\_ML\_PROB2}}:  The percentage likelihood associated specifically with subtypes of {\tt STYPE\_ML} for YSOs and AGB stars, ranging from 0 to 100.  The default value is 0, meaning no machine learning-based classification  has been made at the 80\% confidence level or higher.  
     
 {\bf Column 46. {\tt NAME}}: A commonly used name for the object.  Blank at present (a placeholder for future convenience).

{\bf Column 47. {\tt Gaia\_DR3}}: The {\tt Source} identifier from the {\sl Gaia} DR3 database associated with this target, if available.

{\bf Columns 48, 49. {\tt GaiaDR3\_G}, {\tt GaiaDR3\_Gerr}}:  The {\sl G}-band photometry from {\sl Gaia} DR3 associated with this target, plus the 1$\sigma$ uncertainty, if available.
  
{\bf Columns 50, 51, 52, 53}.  Like columns 46 and 47, but for  {\tt GaiaDR3\_B}, {\tt GaiaDR3\_Berr}, {\tt GaiaDR3\_R}, and {\tt GaiaDR3\_Rerr}, respectively.

{\bf Columns 54, 55, 56, 57, 58, 59}.  Astrometric quantities drawn directly from {\sl Gaia} DR3, if available: 
{\tt Plx}, {\tt e\_Plx},  {\tt pmra}, {\tt pmra\_err}, {\tt pmdec}, and {\tt pmdec\_err}.  {\tt Plx} and {\tt e\_Plx} in particular are the DR3 total parallax and standard error of parallax estimates (in millarcsec) for this source.   

{\bf Columns 60, 61, 62}.  Distance estimates drawn directly from {\sl Gaia} DR3, if available: {\tt Dist}, {\tt e\_Dist\_upper}, {\tt e\_Dist\_lower}.  These are the DR3 distance estimates and uncertainties therein, in parsecs.

{\bf Column 63.  {\tt CLASSvar}}: A string indicating the type of target (if any) associated with the source based on the analysis of its NEOWISE {\sl W1} and {\sl W2} photometry as described in Sec.~\ref{sec:variability}.
See C.\ Contreras et al.\ (2023), in prep, for more details.  The field can be either OAGB or CAGB, indicating a 
significant periodic variability and that the source has colors consistent with those of either an oxygen or carbon AGB star.

{\bf Columns 64, 65.  {\tt W1var}, {\tt W1var\_CLASS}}:  {\tt W1var} is a Boolean variable indicating whether the source was found to be variable based on solely on its NEOWISE {\sl W1} photometry.  {\tt W1var\_CLASS} is
the corresponding variability classification.  This field is either N 
(meaning no variability, if {\tt W1var} is False), or will be set to indicate periodic (P), curved (C), linear (L), burst (B), drop (D), and irregular (I) variability classifications, as defined in \citet{2021ApJ...920..132P}.  Variable sources that do not satisfy any of these \citet{2021ApJ...920..132P} definitions are indicated as unclassified (U).

 {\bf Columns 66, 67.  {\tt W2var}, {\tt W2var\_CLASS}}: As  {\tt W1var} and {\tt W1var\_CLASS} above, but corresponding to the variability and variability class inferred from the NEOWISE {\sl W2} photometry.

{\bf Columns 68 and 69.  {\tt W1VARAVG}, {\tt W2VARAVG}}: The NEOWISE mean magnitudes in the {\sl W1} and {\sl W2} bands, respectively.

{\bf Columns 70 and 71.  {\tt W1VARAMP}, {\tt W2VARAMP}}: The NEOWISE variability amplitudes in the {\sl W1} and {\sl W2} bands, respectively.

The data contained in Columns 63 through 71 are described in more detail in C.\ Contreras et al., (2023), in prep.

    \bigskip

\section{Idiosyncracies and Limitations of the SPHEREx Ices Catalog}\label{sec:illustrations}

SPLICES is based on 2MASS and {\sl WISE} photometry and astrometry.  For that reason some unknown fraction of the targets are spatially unresolved binary stars.  Attempts to fit the spectra of binary stellar systems using single stellar templates will necessarily be inaccurate.

Conversely, users should be aware that the 6\farcs2 isolation criterion also means that SPLICES does not contain
close pairs of attenuated sources.  Despite its size, SPLICES is therefore necessarily incomplete in crowded fields.  

Some sources not in the so-called main survey (e.g., some stars in open stellar clusters) may not have 2MASS PSC magnitudes in SPLICES.  This is because they were selected using different criteria than the primary science targets, without reference to 2MASS photometry and astrometry.

By design SPLICES contains sources brighter than the current best-estimate \SPHEREx\ Band 4-6 bright-source saturation limits.  The reasons are twofold: first, this will in principle allow spectra to be obtained of the continua of  scientifically valuable sources at wavelengths blueward of the onset of saturation even though such spectra are unlikely to contain any strong ice absorption features.  Second, it will prevent confusion on the part of community members who are already familiar with certain very bright ices targets observed by, e.g., {\sl AKARI} and {\sl ISO}, because all such sources satisfying the Ices Investigation selection criteria will be present in SPLICES.

\section{Conclusions}\label{sec:conclusion}

This contribution describes SPLICES, the {\sl SPHEREx} List of Ice Sources, a compilation of nearly nine million spatially isolated, IR-bright point sources likely to exhibit ice absorption features when observed by
NASA's upcoming {\sl SPHEREx} mission.  Because of its well-defined and broad-based selection, based
on bright all-sky broadband photometry, SPLICES targets are accessible to observatories all over the Earth
for followup to characterize e.g., YSOs, protoplanetary disks, and other identified objects (variable sources, carbon stars) in concert with the upcoming multi-epoch {\sl SPHEREx} 0.75-5.0\,$\mu$m spectrophotometry.
Those data, including SPLICES, will be made publicly available for community use at the NASA Infrared Science Archive (IRSA).  
It is hoped that SPLICES, and eventually the forthcoming {\sl SPHEREx} spectrophotometry, will be of great use 
to the community in solving many pressing questions about the origins of organic molecules in stellar systems.

\acknowledgments

Some of the computations carried out in the preparation of this paper were run on the FASRC Cannon cluster 
supported by the FAS Division of Science Research Computing Group at Harvard University.  
The authors are grateful to the Finkbeiner group at the Center for Astrophysics 
for kind assistance with the source-matching; the helpful advice of 
C.\ Zucker and E.\ Schlafly is especially appreciated.
This publication makes use of data products from the {\sl Wide-field Infrared Survey Explorer}, which is a joint project of the University of California, Los Angeles, and the Jet Propulsion Laboratory/California Institute of Technology, funded by the National Aeronautics and Space Administration.
This publication makes use of data products from the Two Micron All Sky Survey, which is a joint project of the University of Massachusetts and the Infrared Processing and Analysis Center/California Institute of Technology, funded by the National Aeronautics and Space Administration and the National Science Foundation.  
This work is based in part on archival data obtained with the {\sl Spitzer Space Telescope}, which was operated by the Jet Propulsion Laboratory, California Institute of Technology under a contract with NASA. 
M.\ K. was supported by Basic Science Research Program
through the National Research Foundation of Korea (NRF)
funded by the Ministry of Science, ICT \& Future Planning
(NRF-2015R1C1A1A01052160). 
MLNA is a Visiting Astronomer at the Infrared Telescope Facility, which is operated by the University of Hawaii under contract 80HQTR19D0030 with the National Aeronautics and Space Administration.
J.-E.\ Lee and C.\ Contreras were supported by the National Research Foundation of Korea 
(NRF) grant funded by the Korea government (MSIT grant number 2021R1A2C1011718).

\vspace{5mm}
\facilities{Infrared Telescope Facility, 2MASS, {\sl WISE}, {\sl Spitzer}(IRAC), {\sl Gaia}}

\software{TOPCAT \citep{2005ASPC..347...29T},  
             {Spextool (Cushing, Vacca, \& Rayner 2014)}
}

		\bibliography{spherex} 

\begin{thebibliography}{}
\expandafter\ifx\csname natexlab\endcsname\relax\def\natexlab#1{#1}\fi
\providecommand{\url}[1]{\href{#1}{#1}}
\providecommand{\dodoi}[1]{doi:~\href{http://doi.org/#1}{\nolinkurl{#1}}}
\providecommand{\doeprint}[1]{\href{http://ascl.net/#1}{\nolinkurl{http://ascl.net/#1}}}
\providecommand{\doarXiv}[1]{\href{https://arxiv.org/abs/#1}{\nolinkurl{https://arxiv.org/abs/#1}}}

\bibitem[{{Agarwal} {et~al.}(2021){Agarwal}, {Rao}, {Vaidya}, \&
  {Bhattacharya}}]{2021MNRAS.502.2582A}
{Agarwal}, M., {Rao}, K.~K., {Vaidya}, K., \& {Bhattacharya}, S. 2021, \mnras,
  502, 2582, \dodoi{10.1093/mnras/stab118}

\bibitem[{{Aikawa} {et~al.}(2012){Aikawa}, {Kamuro}, {Sakon}, {Itoh}, {Terada},
  {Noble}, {Pontoppidan}, {Fraser}, {Tamura}, {Kandori}, {Kawamura}, \&
  {Ueno}}]{2012A&A...538A..57A}
{Aikawa}, Y., {Kamuro}, D., {Sakon}, I., {et~al.} 2012, \aap, 538, A57,
  \dodoi{10.1051/0004-6361/201015999}

\bibitem[{{Akeson} {et~al.}(2019){Akeson}, {Jensen}, {Carpenter}, {Ricci},
  {Laos}, {Nogueira}, \& {Suen-Lewis}}]{2019ApJ...872..158A}
{Akeson}, R.~L., {Jensen}, E. L.~N., {Carpenter}, J., {et~al.} 2019, \apj, 872,
  158, \dodoi{10.3847/1538-4357/aaff6a}

\bibitem[{{Allen} {et~al.}(2012){Allen}, {Gutermuth}, {Kryukova}, {Megeath},
  {Pipher}, {Naylor}, {Jeffries}, {Wolk}, {Spitzbart}, \&
  {Muzerolle}}]{Allen.2012}
{Allen}, T.~S., {Gutermuth}, R.~A., {Kryukova}, E., {et~al.} 2012, \apj, 750,
  125, \dodoi{10.1088/0004-637X/750/2/125}

\bibitem[{{Andrews} {et~al.}(2018){Andrews}, {Huang}, {P{\'e}rez}, {Isella},
  {Dullemond}, {Kurtovic}, {Guzm{\'a}n}, {Carpenter}, {Wilner}, {Zhang}, {Zhu},
  {Birnstiel}, {Bai}, {Benisty}, {Hughes}, {{\"O}berg}, \&
  {Ricci}}]{2018ApJ...869L..41A}
{Andrews}, S.~M., {Huang}, J., {P{\'e}rez}, L.~M., {et~al.} 2018, \apjl, 869,
  L41, \dodoi{10.3847/2041-8213/aaf741}

\bibitem[{{Azimlu} {et~al.}(2015){Azimlu}, {Mart{\'\i}nez-Galarza}, \&
  {Muench}}]{Azimlu.2015}
{Azimlu}, M., {Mart{\'\i}nez-Galarza}, J.~R., \& {Muench}, A.~A. 2015, \aj,
  150, 95, \dodoi{10.1088/0004-6256/150/3/95}

\bibitem[{{Beerer} {et~al.}(2010){Beerer}, {Koenig}, {Hora}, {Gutermuth},
  {Bontemps}, {Megeath}, {Schneider}, {Motte}, {Carey}, {Simon}, {Keto},
  {Smith}, {Allen}, {Fazio}, {Kraemer}, {Price}, {Mizuno}, {Adams},
  {Hern{\'a}ndez}, \& {Lucas}}]{2010ApJ...720..679B}
{Beerer}, I.~M., {Koenig}, X.~P., {Hora}, J.~L., {et~al.} 2010, \apj, 720, 679,
  \dodoi{10.1088/0004-637X/720/1/679}

\bibitem[{{Billot} {et~al.}(2010){Billot}, {Noriega-Crespo}, {Carey}, {Guieu},
  {Shenoy}, {Paladini}, \& {Latter}}]{Billot.2010}
{Billot}, N., {Noriega-Crespo}, A., {Carey}, S., {et~al.} 2010, \apj, 712, 797,
  \dodoi{10.1088/0004-637X/712/2/797}

\bibitem[{{Boogert} {et~al.}(2015){Boogert}, {Gerakines}, \&
  {Whittet}}]{2015ARA&A..53..541B}
{Boogert}, A.~C.~A., {Gerakines}, P.~A., \& {Whittet}, D. C.~B. 2015, \araa,
  53, 541, \dodoi{10.1146/annurev-astro-082214-122348}

\bibitem[{{Brown} {et~al.}(2013){Brown}, {Pontoppidan}, {van Dishoeck},
  {Herczeg}, {Blake}, \& {Smette}}]{2013ApJ...770...94B}
{Brown}, J.~M., {Pontoppidan}, K.~M., {van Dishoeck}, E.~F., {et~al.} 2013,
  \apj, 770, 94, \dodoi{10.1088/0004-637X/770/2/94}

\bibitem[{{Cambr{\'e}sy} {et~al.}(2013){Cambr{\'e}sy}, {Marton}, {Feher},
  {T{\'o}th}, \& {Schneider}}]{Cambresy.2013}
{Cambr{\'e}sy}, L., {Marton}, G., {Feher}, O., {T{\'o}th}, L.~V., \&
  {Schneider}, N. 2013, \aap, 557, A29, \dodoi{10.1051/0004-6361/201321235}

\bibitem[{{Cantat-Gaudin} {et~al.}(2018){Cantat-Gaudin}, {Vallenari}, {Sordo},
  {Pensabene}, {Krone-Martins}, {Moitinho}, {Jordi}, {Casamiquela},
  {Balaguer-N{\'u}nez}, {Soubiran}, \& {Brouillet}}]{2018A&A...615A..49C}
{Cantat-Gaudin}, T., {Vallenari}, A., {Sordo}, R., {et~al.} 2018, \aap, 615,
  A49, \dodoi{10.1051/0004-6361/201731251}

\bibitem[{{Carlson} {et~al.}(2012){Carlson}, {Sewi{\l}o}, {Meixner}, {Romita},
  \& {Lawton}}]{Carlson.2012}
{Carlson}, L.~R., {Sewi{\l}o}, M., {Meixner}, M., {Romita}, K.~A., \& {Lawton},
  B. 2012, \aap, 542, A66, \dodoi{10.1051/0004-6361/201118627}

\bibitem[{{Chambers} {et~al.}(2016){Chambers}, {Magnier}, {Metcalfe},
  {Flewelling}, {Huber}, {Waters}, {Denneau}, {Draper}, {Farrow}, {Finkbeiner},
  {Holmberg}, {Koppenhoefer}, {Price}, {Rest}, {Saglia}, {Schlafly}, {Smartt},
  {Sweeney}, {Wainscoat}, {Burgett}, {Chastel}, {Grav}, {Heasley}, {Hodapp},
  {Jedicke}, {Kaiser}, {Kudritzki}, {Luppino}, {Lupton}, {Monet}, {Morgan},
  {Onaka}, {Shiao}, {Stubbs}, {Tonry}, {White}, {Ba{\~n}ados}, {Bell},
  {Bender}, {Bernard}, {Boegner}, {Boffi}, {Botticella}, {Calamida},
  {Casertano}, {Chen}, {Chen}, {Cole}, {Deacon}, {Frenk}, {Fitzsimmons},
  {Gezari}, {Gibbs}, {Goessl}, {Goggia}, {Gourgue}, {Goldman}, {Grant},
  {Grebel}, {Hambly}, {Hasinger}, {Heavens}, {Heckman}, {Henderson}, {Henning},
  {Holman}, {Hopp}, {Ip}, {Isani}, {Jackson}, {Keyes}, {Koekemoer}, {Kotak},
  {Le}, {Liska}, {Long}, {Lucey}, {Liu}, {Martin}, {Masci}, {McLean}, {Mindel},
  {Misra}, {Morganson}, {Murphy}, {Obaika}, {Narayan}, {Nieto-Santisteban},
  {Norberg}, {Peacock}, {Pier}, {Postman}, {Primak}, {Rae}, {Rai}, {Riess},
  {Riffeser}, {Rix}, {R{\"o}ser}, {Russel}, {Rutz}, {Schilbach}, {Schultz},
  {Scolnic}, {Strolger}, {Szalay}, {Seitz}, {Small}, {Smith}, {Soderblom},
  {Taylor}, {Thomson}, {Taylor}, {Thakar}, {Thiel}, {Thilker}, {Unger},
  {Urata}, {Valenti}, {Wagner}, {Walder}, {Walter}, {Watters}, {Werner},
  {Wood-Vasey}, \& {Wyse}}]{2016arXiv161205560C}
{Chambers}, K.~C., {Magnier}, E.~A., {Metcalfe}, N., {et~al.} 2016, arXiv
  e-prints, arXiv:1612.05560.
\newblock \doarXiv{1612.05560}

\bibitem[{{Chavarr{\'\i}a} {et~al.}(2014){Chavarr{\'\i}a}, {Allen}, {Brunt},
  {Hora}, {Muench}, \& {Fazio}}]{Chavarria.2014}
{Chavarr{\'\i}a}, L., {Allen}, L., {Brunt}, C., {et~al.} 2014, \mnras, 439,
  3719, \dodoi{10.1093/mnras/stu224}

\bibitem[{{Churchwell} {et~al.}(2009){Churchwell}, {Babler}, {Meade},
  {Whitney}, {Benjamin}, {Indebetouw}, {Cyganowski}, {Robitaille}, {Povich},
  {Watson}, \& {Bracker}}]{2009PASP..121..213C}
{Churchwell}, E., {Babler}, B.~L., {Meade}, M.~R., {et~al.} 2009, \pasp, 121,
  213, \dodoi{10.1086/597811}

\bibitem[{{Cohen} {et~al.}(1988){Cohen}, {Dame}, {Garay}, {Montani}, {Rubio},
  \& {Thaddeus}}]{1988ApJ...331L..95C}
{Cohen}, R.~S., {Dame}, T.~M., {Garay}, G., {et~al.} 1988, \apjl, 331, L95,
  \dodoi{10.1086/185243}

\bibitem[{{Connelley} {et~al.}(2008){Connelley}, {Reipurth}, \&
  {Tokunaga}}]{Connelley.2008}
{Connelley}, M.~S., {Reipurth}, B., \& {Tokunaga}, A.~T. 2008, \aj, 135, 2496,
  \dodoi{10.1088/0004-6256/135/6/2496}

\bibitem[{{Cottle} {et~al.}(2018){Cottle}, {Covey}, {Su{\'a}rez},
  {Rom{\'a}n-Z{\'u}{\~n}iga}, {Schlafly}, {Downes}, {Ybarra}, {Hernandez},
  {Stassun}, {Stringfellow}, {Getman}, {Feigelson}, {Borissova}, {Kim},
  {Roman-Lopes}, {Da Rio}, {De Lee}, {Frinchaboy}, {Kounkel}, {Majewski},
  {Mennickent}, {Nidever}, {Nitschelm}, {Pan}, {Shetrone}, {Zasowski},
  {Chambers}, {Magnier}, \& {Valenti}}]{Cottle.2018}
{Cottle}, J.~N., {Covey}, K.~R., {Su{\'a}rez}, G., {et~al.} 2018, \apjs, 236,
  27, \dodoi{10.3847/1538-4365/aabada}

\bibitem[{{Crill} {et~al.}(2020){Crill}, {Werner}, {Akeson}, {Ashby}, {Bleem},
  {Bock}, {Bryan}, {Burnham}, {Byunh}, {Chang}, {Chiang}, {Cook}, {Cooray},
  {Davis}, {Dor{\'e}}, {Dowell}, {Dubois-Felsmann}, {Eifler}, {Faisst},
  {Habib}, {Heinrich}, {Heitmann}, {Heaton}, {Hirata}, {Hristov}, {Hui},
  {Jeong}, {Kang}, {Kecman}, {Kirkpatrick}, {Korngut}, {Krause}, {Lee},
  {Lisse}, {Masters}, {Mauskopf}, {Melnick}, {Miyasaka}, {Nayyeri}, {Nguyen},
  {{\"O}berg}, {Padin}, {Paladini}, {Pourrahmani}, {Pyo}, {Smith}, {Song},
  {Symons}, {Teplitz}, {Tolls}, {Unwin}, {Windhorst}, {Yang}, \&
  {Zemcov}}]{Crill.2020}
{Crill}, B.~P., {Werner}, M., {Akeson}, R., {et~al.} 2020, in Society of
  Photo-Optical Instrumentation Engineers (SPIE) Conference Series, Vol. 11443,
  Society of Photo-Optical Instrumentation Engineers (SPIE) Conference Series,
  114430I, \dodoi{10.1117/12.2567224}

\bibitem[{{Cutri} {et~al.}(2021){Cutri}, {Wright}, {Conrow}, {Fowler},
  {Eisenhardt}, {Grillmair}, {Kirkpatrick}, {Masci}, {McCallon}, {Wheelock},
  {Fajardo-Acosta}, {Yan}, {Benford}, {Harbut}, {Jarrett}, {Lake}, {Leisawitz},
  {Ressler}, {Stanford}, {Tsai}, {Liu}, {Helou}, {Mainzer}, {Gettngs},
  {Gonzalez}, {Hoffman}, {Marsh}, {Padgett}, {Skrutskie}, {Beck}, {Papin}, \&
  {Wittman}}]{2014yCat.2328....0C}
{Cutri}, R.~M., {Wright}, E.~L., {Conrow}, T., {et~al.} 2021, VizieR Online
  Data Catalog, II/328

\bibitem[{{Dame} {et~al.}(2001){Dame}, {Hartmann}, \&
  {Thaddeus}}]{2001ApJ...547..792D}
{Dame}, T.~M., {Hartmann}, D., \& {Thaddeus}, P. 2001, \apj, 547, 792,
  \dodoi{10.1086/318388}

\bibitem[{{Dent} {et~al.}(2013){Dent}, {Thi}, {Kamp}, {Williams}, {Menard},
  {Andrews}, {Ardila}, {Aresu}, {Augereau}, {Barrado y Navascues}, {Brittain},
  {Carmona}, {Ciardi}, {Danchi}, {Donaldson}, {Duchene}, {Eiroa}, {Fedele},
  {Grady}, {de Gregorio-Molsalvo}, {Howard}, {Hu{\'e}lamo}, {Krivov},
  {Lebreton}, {Liseau}, {Martin-Zaidi}, {Mathews}, {Meeus}, {Mendigut{\'\i}a},
  {Montesinos}, {Morales-Calderon}, {Mora}, {Nomura}, {Pantin}, {Pascucci},
  {Phillips}, {Pinte}, {Podio}, {Ramsay}, {Riaz}, {Riviere-Marichalar},
  {Roberge}, {Sandell}, {Solano}, {Tilling}, {Torrelles}, {Vandenbusche},
  {Vicente}, {White}, \& {Woitke}}]{Dent.2013}
{Dent}, W.~R.~F., {Thi}, W.~F., {Kamp}, I., {et~al.} 2013, \pasp, 125, 477,
  \dodoi{10.1086/670826}

\bibitem[{{Dunham} {et~al.}(2015){Dunham}, {Allen}, {Evans},
  {Broekhoven-Fiene}, {Cieza}, {Di Francesco}, {Gutermuth}, {Harvey},
  {Hatchell}, {Heiderman}, {Huard}, {Johnstone}, {Kirk}, {Matthews}, {Miller},
  {Peterson}, \& {Young}}]{Dunham.2015}
{Dunham}, M.~M., {Allen}, L.~E., {Evans}, Neal~J., I., {et~al.} 2015, \apjs,
  220, 11, \dodoi{10.1088/0067-0049/220/1/11}

\bibitem[{{Esplin} \& {Luhman}(2019)}]{Esplin.2019}
{Esplin}, T.~L., \& {Luhman}, K.~L. 2019, \aj, 158, 54,
  \dodoi{10.3847/1538-3881/ab2594}

\bibitem[{{Fang} {et~al.}(2013){Fang}, {Kim}, {van Boekel}, {Sicilia-Aguilar},
  {Henning}, \& {Flaherty}}]{Fang.2013}
{Fang}, M., {Kim}, J.~S., {van Boekel}, R., {et~al.} 2013, \apjs, 207, 5,
  \dodoi{10.1088/0067-0049/207/1/5}

\bibitem[{{Fazio} {et~al.}(2004){Fazio}, {Hora}, {Allen}, {Ashby}, {Barmby},
  {Deutsch}, {Huang}, {Kleiner}, {Marengo}, {Megeath}, {Melnick}, {Pahre},
  {Patten}, {Polizotti}, {Smith}, {Taylor}, {Wang}, {Willner}, {Hoffmann},
  {Pipher}, {Forrest}, {McMurty}, {McCreight}, {McKelvey}, {McMurray}, {Koch},
  {Moseley}, {Arendt}, {Mentzell}, {Marx}, {Losch}, {Mayman}, {Eichhorn},
  {Krebs}, {Jhabvala}, {Gezari}, {Fixsen}, {Flores}, {Shakoorzadeh}, {Jungo},
  {Hakun}, {Workman}, {Karpati}, {Kichak}, {Whitley}, {Mann}, {Tollestrup},
  {Eisenhardt}, {Stern}, {Gorjian}, {Bhattacharya}, {Carey}, {Nelson},
  {Glaccum}, {Lacy}, {Lowrance}, {Laine}, {Reach}, {Stauffer}, {Surace},
  {Wilson}, {Wright}, {Hoffman}, {Domingo}, \& {Cohen}}]{2004ApJS..154...10F}
{Fazio}, G.~G., {Hora}, J.~L., {Allen}, L.~E., {et~al.} 2004, \apjs, 154, 10,
  \dodoi{10.1086/422843}

\bibitem[{{Fischer} {et~al.}(2016){Fischer}, {Padgett}, {Stapelfeldt}, \&
  {Sewi{\l}o}}]{Fischer.2016}
{Fischer}, W.~J., {Padgett}, D.~L., {Stapelfeldt}, K.~L., \& {Sewi{\l}o}, M.
  2016, \apj, 827, 96, \dodoi{10.3847/0004-637X/827/2/96}

\bibitem[{{Flaherty} {et~al.}(2013){Flaherty}, {Muzerolle}, {Rieke},
  {Gutermuth}, {Balog}, {Herbst}, \& {Megeath}}]{Flaherty.2013}
{Flaherty}, K.~M., {Muzerolle}, J., {Rieke}, G., {et~al.} 2013, \aj, 145, 66,
  \dodoi{10.1088/0004-6256/145/3/66}

\bibitem[{{Gaia Collaboration} {et~al.}(2023){Gaia Collaboration}, {Vallenari},
  {Brown}, {Prusti}, {de Bruijne}, {Arenou}, {Babusiaux}, {Biermann},
  {Creevey}, {Ducourant}, {Evans}, {Eyer}, {Guerra}, {Hutton}, {Jordi},
  {Klioner}, {Lammers}, {Lindegren}, {Luri}, {Mignard}, {Panem}, {Pourbaix},
  {Randich}, {Sartoretti}, {Soubiran}, {Tanga}, {Walton}, {Bailer-Jones},
  {Bastian}, {Drimmel}, {Jansen}, {Katz}, {Lattanzi}, {van Leeuwen}, {Bakker},
  {Cacciari}, {Casta{\~n}eda}, {De Angeli}, {Fabricius}, {Fouesneau},
  {Fr{\'e}mat}, {Galluccio}, {Guerrier}, {Heiter}, {Masana}, {Messineo},
  {Mowlavi}, {Nicolas}, {Nienartowicz}, {Pailler}, {Panuzzo}, {Riclet}, {Roux},
  {Seabroke}, {Sordo{\o}rcit}, {Th{\'e}venin}, {Gracia-Abril}, {Portell},
  {Teyssier}, {Altmann}, {Andrae}, {Audard}, {Bellas-Velidis}, {Benson},
  {Berthier}, {Blomme}, {Burgess}, {Busonero}, {Busso}, {C{\'a}novas}, {Carry},
  {Cellino}, {Cheek}, {Clementini}, {Damerdji}, {Davidson}, {de Teodoro},
  {Nu{\~n}ez Campos}, {Delchambre}, {Dell'Oro}, {Esquej},
  {Fern{\'a}ndez-Hern{\'a}ndez}, {Fraile}, {Garabato}, {Garc{\'\i}a-Lario},
  {Gosset}, {Haigron}, {Halbwachs}, {Hambly}, {Harrison}, {Hern{\'a}ndez},
  {Hestroffer}, {Hodgkin}, {Holl}, {Jan{\ss}en}, {Jevardat de Fombelle},
  {Jordan}, {Krone-Martins}, {Lanzafame}, {L{\"o}ffler}, {Marchal}, {Marrese},
  {Moitinho}, {Muinonen}, {Osborne}, {Pancino}, {Pauwels}, {Recio-Blanco},
  {Reyl{\'e}}, {Riello}, {Rimoldini}, {Roegiers}, {Rybizki}, {Sarro}, {Siopis},
  {Smith}, {Sozzetti}, {Utrilla}, {van Leeuwen}, {Abbas}, {{\'A}brah{\'a}m},
  {Abreu Aramburu}, {Aerts}, {Aguado}, {Ajaj}, {Aldea-Montero}, {Altavilla},
  {{\'A}lvarez}, {Alves}, {Anders}, {Anderson}, {Anglada Varela}, {Antoja},
  {Baines}, {Baker}, {Balaguer-N{\'u}{\~n}ez}, {Balbinot}, {Balog}, {Barache},
  {Barbato}, {Barros}, {Barstow}, {Bartolom{\'e}}, {Bassilana}, {Bauchet},
  {Becciani}, {Bellazzini}, {Berihuete}, {Bernet}, {Bertone}, {Bianchi},
  {Binnenfeld}, {Blanco-Cuaresma}, {Blazere}, {Boch}, {Bombrun}, {Bossini},
  {Bouquillon}, {Bragaglia}, {Bramante}, {Breedt}, {Bressan}, {Brouillet},
  {Brugaletta}, {Bucciarelli}, {Burlacu}, {Butkevich}, {Buzzi}, {Caffau},
  {Cancelliere}, {Cantat-Gaudin}, {Carballo}, {Carlucci}, {Carnerero},
  {Carrasco}, {Casamiquela}, {Castellani}, {Castro-Ginard}, {Chaoul},
  {Charlot}, {Chemin}, {Chiaramida}, {Chiavassa}, {Chornay}, {Comoretto},
  {Contursi}, {Cooper}, {Cornez}, {Cowell}, {Crifo}, {Cropper}, {Crosta},
  {Crowley}, {Dafonte}, {Dapergolas}, {David}, {David}, {de Laverny}, {De
  Luise}, {De March}, {De Ridder}, {de Souza}, {de Torres}, {del Peloso}, {del
  Pozo}, {Delbo}, {Delgado}, {Delisle}, {Demouchy}, {Dharmawardena}, {Di
  Matteo}, {Diakite}, {Diener}, {Distefano}, {Dolding}, {Edvardsson}, {Enke},
  {Fabre}, {Fabrizio}, {Faigler}, {Fedorets}, {Fernique}, {Fienga}, {Figueras},
  {Fournier}, {Fouron}, {Fragkoudi}, {Gai}, {Garcia-Gutierrez},
  {Garcia-Reinaldos}, {Garc{\'\i}a-Torres}, {Garofalo}, {Gavel}, {Gavras},
  {Gerlach}, {Geyer}, {Giacobbe}, {Gilmore}, {Girona}, {Giuffrida}, {Gomel},
  {Gomez}, {Gonz{\'a}lez-N{\'u}{\~n}ez}, {Gonz{\'a}lez-Santamar{\'\i}a},
  {Gonz{\'a}lez-Vidal}, {Granvik}, {Guillout}, {Guiraud},
  {Guti{\'e}rrez-S{\'a}nchez}, {Guy}, {Hatzidimitriou}, {Hauser}, {Haywood},
  {Helmer}, {Helmi}, {Sarmiento}, {Hidalgo}, {Hilger}, {H{\l}adczuk}, {Hobbs},
  {Holland}, {Huckle}, {Jardine}, {Jasniewicz}, {Jean-Antoine Piccolo},
  {Jim{\'e}nez-Arranz}, {Jorissen}, {Juaristi Campillo}, {Julbe}, {Karbevska},
  {Kervella}, {Khanna}, {Kontizas}, {Kordopatis}, {Korn}, {K{\'o}sp{\'a}l},
  {Kostrzewa-Rutkowska}, {Kruszy{\'n}ska}, {Kun}, {Laizeau}, {Lambert},
  {Lanza}, {Lasne}, {Le Campion}, {Lebreton}, {Lebzelter}, {Leccia}, {Leclerc},
  {Lecoeur-Taibi}, {Liao}, {Licata}, {Lindstr{\o}m}, {Lister}, {Livanou},
  {Lobel}, {Lorca}, {Loup}, {Madrero Pardo}, {Magdaleno Romeo}, {Managau},
  {Mann}, {Manteiga}, {Marchant}, {Marconi}, {Marcos}, {Marcos Santos},
  {Mar{\'\i}n Pina}, {Marinoni}, {Marocco}, {Marshall}, {Polo},
  {Mart{\'\i}n-Fleitas}, {Marton}, {Mary}, {Masip}, {Massari},
  {Mastrobuono-Battisti}, {Mazeh}, {McMillan}, {Messina}, {Michalik}, {Millar},
  {Mints}, {Molina}, {Molinaro}, {Moln{\'a}r}, {Monari}, {Mongui{\'o}},
  {Montegriffo}, {Montero}, {Mor}, {Mora}, {Morbidelli}, {Morel}, {Morris},
  {Muraveva}, {Murphy}, {Musella}, {Nagy}, {Noval}, {Oca{\~n}a}, {Ogden},
  {Ordenovic}, {Osinde}, {Pagani}, {Pagano}, {Palaversa}, {Palicio},
  {Pallas-Quintela}, {Panahi}, {Payne-Wardenaar}, {Pe{\~n}alosa Esteller},
  {Penttil{\"a}}, {Pichon}, {Piersimoni}, {Pineau}, {Plachy}, {Plum}, {Poggio},
  {Pr{\v{s}}a}, {Pulone}, {Racero}, {Ragaini}, {Rainer}, {Raiteri}, {Rambaux},
  {Ramos}, {Ramos-Lerate}, {Re Fiorentin}, {Regibo}, {Richards}, {Rios Diaz},
  {Ripepi}, {Riva}, {Rix}, {Rixon}, {Robichon}, {Robin}, {Robin}, {Roelens},
  {Rogues}, {Rohrbasser}, {Romero-G{\'o}mez}, {Rowell}, {Royer}, {Ruz Mieres},
  {Rybicki}, {Sadowski}, {S{\'a}ez N{\'u}{\~n}ez}, {Sagrist{\`a} Sell{\'e}s},
  {Sahlmann}, {Salguero}, {Samaras}, {Sanchez Gimenez}, {Sanna},
  {Santove{\~n}a}, {Sarasso}, {Schultheis}, {Sciacca}, {Segol}, {Segovia},
  {S{\'e}gransan}, {Semeux}, {Shahaf}, {Siddiqui}, {Siebert}, {Siltala},
  {Silvelo}, {Slezak}, {Slezak}, {Smart}, {Snaith}, {Solano}, {Solitro},
  {Souami}, {Souchay}, {Spagna}, {Spina}, {Spoto}, {Steele},
  {Steidelm{\"u}ller}, {Stephenson}, {S{\"u}veges}, {Surdej}, {Szabados},
  {Szegedi-Elek}, {Taris}, {Taylo}, {Teixeira}, {Tolomei}, {Tonello}, {Torra},
  {Torra}, {Torralba Elipe}, {Trabucchi}, {Tsounis}, {Turon}, {Ulla}, {Unger},
  {Vaillant}, {van Dillen}, {van Reeven}, {Vanel}, {Vecchiato}, {Viala},
  {Vicente}, {Voutsinas}, {Weiler}, {Wevers}, {Wyrzykowski}, {Yoldas}, {Yvard},
  {Zhao}, {Zorec}, {Zucker}, \& {Zwitter}}]{2022arXiv220800211G}
{Gaia Collaboration}, {Vallenari}, A., {Brown}, A.~G.~A., {et~al.} 2023, \aap,
  674, A1, \dodoi{10.1051/0004-6361/202243940}

\bibitem[{{Gama} {et~al.}(2016){Gama}, {Lepine}, {Mendoza}, {Wu}, \&
  {Yuan}}]{Gama.2016}
{Gama}, D.~R.~G., {Lepine}, J.~R.~D., {Mendoza}, E., {Wu}, Y., \& {Yuan}, J.
  2016, \apj, 830, 57, \dodoi{10.3847/0004-637X/830/2/57}

\bibitem[{{Gibb} {et~al.}(2004){Gibb}, {Whittet}, {Boogert}, \&
  {Tielens}}]{2004ApJS..151...35G}
{Gibb}, E.~L., {Whittet}, D.~C.~B., {Boogert}, A.~C.~A., \& {Tielens},
  A.~G.~G.~M. 2004, \apjs, 151, 35, \dodoi{10.1086/381182}

\bibitem[{{Gordon} {et~al.}(2011){Gordon}, {Meixner}, {Meade}, {Whitney},
  {Engelbracht}, {Bot}, {Boyer}, {Lawton}, {Sewi{\l}o}, {Babler}, {Bernard},
  {Bracker}, {Block}, {Blum}, {Bolatto}, {Bonanos}, {Harris}, {Hora},
  {Indebetouw}, {Misselt}, {Reach}, {Shiao}, {Tielens}, {Carlson},
  {Churchwell}, {Clayton}, {Chen}, {Cohen}, {Fukui}, {Gorjian}, {Hony},
  {Israel}, {Kawamura}, {Kemper}, {Leroy}, {Li}, {Madden}, {Marble},
  {McDonald}, {Mizuno}, {Mizuno}, {Muller}, {Oliveira}, {Olsen}, {Onishi},
  {Paladini}, {Paradis}, {Points}, {Robitaille}, {Rubin}, {Sandstrom}, {Sato},
  {Shibai}, {Simon}, {Smith}, {Srinivasan}, {Vijh}, {Van Dyk}, {van Loon}, \&
  {Zaritsky}}]{2011AJ....142..102G}
{Gordon}, K.~D., {Meixner}, M., {Meade}, M.~R., {et~al.} 2011, \aj, 142, 102,
  \dodoi{10.1088/0004-6256/142/4/102}

\bibitem[{{Greene} {et~al.}(1994){Greene}, {Wilking}, {Andre}, {Young}, \&
  {Lada}}]{1994ApJ...434..614G}
{Greene}, T.~P., {Wilking}, B.~A., {Andre}, P., {Young}, E.~T., \& {Lada},
  C.~J. 1994, \apj, 434, 614, \dodoi{10.1086/174763}

\bibitem[{{Gro{\ss}schedl} {et~al.}(2019){Gro{\ss}schedl}, {Alves}, {Teixeira},
  {Bouy}, {Forbrich}, {Lada}, {Meingast}, {Hacar}, {Ascenso}, {Ackerl},
  {Hasenberger}, {K{\"o}hler}, {Kubiak}, {Larreina}, {Linhardt}, {Lombardi}, \&
  {M{\"o}ller}}]{Grossschedl.2019}
{Gro{\ss}schedl}, J.~E., {Alves}, J., {Teixeira}, P.~S., {et~al.} 2019, \aap,
  622, A149, \dodoi{10.1051/0004-6361/201832577}

\bibitem[{{Gutermuth} {et~al.}(2009){Gutermuth}, {Megeath}, {Myers}, {Allen},
  {Pipher}, \& {Fazio}}]{Gutermuth.2009}
{Gutermuth}, R.~A., {Megeath}, S.~T., {Myers}, P.~C., {et~al.} 2009, \apjs,
  184, 18, \dodoi{10.1088/0067-0049/184/1/18}

\bibitem[{{Heiter} {et~al.}(2015){Heiter}, {Jofr{\'e}}, {Gustafsson}, {Korn},
  {Soubiran}, \& {Th{\'e}venin}}]{2015A&A...582A..49H}
{Heiter}, U., {Jofr{\'e}}, P., {Gustafsson}, B., {et~al.} 2015, \aap, 582, A49,
  \dodoi{10.1051/0004-6361/201526319}

\bibitem[{{Heyer} \& {Dame}(2015)}]{2015ARA&A..53..583H}
{Heyer}, M., \& {Dame}, T.~M. 2015, \araa, 53, 583,
  \dodoi{10.1146/annurev-astro-082214-122324}

\bibitem[{{Hollenbach} {et~al.}(2009){Hollenbach}, {Kaufman}, {Bergin}, \&
  {Melnick}}]{2009ApJ...690.1497H}
{Hollenbach}, D., {Kaufman}, M.~J., {Bergin}, E.~A., \& {Melnick}, G.~J. 2009,
  \apj, 690, 1497, \dodoi{10.1088/0004-637X/690/2/1497}

\bibitem[{Hora {et~al.}(2021)Hora, Bontemps, Megeath, Schneider, Motte,
  Gutermuth, Carey, Mizuno, Kraemer, Simon, Keto, Smith, Allen, Fazio, Adams,
  Price, \& Koenig}]{DVN/1QMXB8_2021}
Hora, J.~L., Bontemps, S., Megeath, S.~T., {et~al.} 2021, {Spitzer Cygnus-X
  Legacy Project}, V2,  Harvard Dataverse, \dodoi{10.7910/DVN/1QMXB8}

\bibitem[{{Ishii} {et~al.}(1998){Ishii}, {Nagata}, {Sato}, {Watanabe}, {Yao},
  \& {Jones}}]{1998AJ....116..868I}
{Ishii}, M., {Nagata}, T., {Sato}, S., {et~al.} 1998, \aj, 116, 868,
  \dodoi{10.1086/300467}

\bibitem[{{Kim} {et~al.}(2015){Kim}, {Koo}, \& {Davis}}]{Kim.2015}
{Kim}, H.-J., {Koo}, B.-C., \& {Davis}, C.~J. 2015, \apj, 802, 59,
  \dodoi{10.1088/0004-637X/802/1/59}

\bibitem[{{Kim} {et~al.}(2022){Kim}, {Lee}, {Jeong}, {Kim}, {Aikawa}, {Noble},
  {Choi}, {Lee}, {Dunham}, {Kim}, \& {Koo}}]{2022ApJ...935..137K}
{Kim}, J., {Lee}, J.-E., {Jeong}, W.-S., {et~al.} 2022, \apj, 935, 137,
  \dodoi{10.3847/1538-4357/ac7f9f}

\bibitem[{{Koenig} {et~al.}(2015){Koenig}, {Hillenbrand}, {Padgett}, \&
  {DeFelippis}}]{Koenig.2015}
{Koenig}, X., {Hillenbrand}, L.~A., {Padgett}, D.~L., \& {DeFelippis}, D. 2015,
  \aj, 150, 100, \dodoi{10.1088/0004-6256/150/4/100}

\bibitem[{{Koenig} {et~al.}(2008){Koenig}, {Allen}, {Gutermuth}, {Hora},
  {Brunt}, \& {Muzerolle}}]{Koenig.2008}
{Koenig}, X.~P., {Allen}, L.~E., {Gutermuth}, R.~A., {et~al.} 2008, \apj, 688,
  1142, \dodoi{10.1086/592322}

\bibitem[{{Koenig} \& {Leisawitz}(2014)}]{2014Koenig}
{Koenig}, X.~P., \& {Leisawitz}, D.~T. 2014, \apj, 791, 131,
  \dodoi{10.1088/0004-637X/791/2/131}

\bibitem[{{Korngut} {et~al.}(2018){Korngut}, {Bock}, {Akeson}, {Ashby},
  {Bleem}, {Boland}, {Bolton}, {Bradford}, {Braun}, {Bryan}, {Capak}, {Chang},
  {Coffey}, {Cooray}, {Crill}, {Dor{\'e}}, {Eifler}, {Feng}, {Habib},
  {Heitmann}, {Hemmati}, {Hirata}, {Jeong}, {Kim}, {Kirkpatrick},
  {Kowalkowski}, {Krause}, {Lisse}, {Mauskopf}, {Masters}, {McGuire},
  {Melnick}, {Nguyen}, {Nayyeri}, {Oberg}, {dePutter}, {Purcell}, {Rocca},
  {Runyan}, {Sandstrom}, {Smith}, {Song}, {Stickley}, {Stober}, {Susca},
  {Teplitz}, {Tolls}, {Unwin}, {Werner}, {Windhorst}, \&
  {Zemcov}}]{2018SPIE10698E..1UK}
{Korngut}, P.~M., {Bock}, J.~J., {Akeson}, R., {et~al.} 2018, in Society of
  Photo-Optical Instrumentation Engineers (SPIE) Conference Series, Vol. 10698,
  Space Telescopes and Instrumentation 2018: Optical, Infrared, and Millimeter
  Wave, ed. M.~{Lystrup}, H.~A. {MacEwen}, G.~G. {Fazio}, N.~{Batalha},
  N.~{Siegler}, \& E.~C. {Tong}, 106981U, \dodoi{10.1117/12.2312860}

\bibitem[{{Kounkel} {et~al.}(2014){Kounkel}, {Hartmann}, {Loinard},
  {Mioduszewski}, {Dzib}, {Ortiz-Le{\'o}n}, {Rodr{\'\i}guez}, {Pech}, {Rivera},
  {Torres}, {Boden}, {Evans}, {Brice{\~n}o}, \& {Tobin}}]{Kounkel.2014}
{Kounkel}, M., {Hartmann}, L., {Loinard}, L., {et~al.} 2014, \apj, 790, 49,
  \dodoi{10.1088/0004-637X/790/1/49}

\bibitem[{{Kuhn} {et~al.}(2021){Kuhn}, {de Souza}, {Krone-Martins},
  {Castro-Ginard}, {Ishida}, {Povich}, {Hillenbrand}, \& {COIN
  Collaboration}}]{Kuhn.2021}
{Kuhn}, M.~A., {de Souza}, R.~S., {Krone-Martins}, A., {et~al.} 2021, \apjs,
  254, 33, \dodoi{10.3847/1538-4365/abe465}

\bibitem[{{Kun} {et~al.}(2016){Kun}, {Wolf-Chase}, {Mo{\'o}r}, {Apai}, {Balog},
  {O'Linger-Luscusk}, \& {Moriarty-Schieven}}]{Kun.2016}
{Kun}, M., {Wolf-Chase}, G., {Mo{\'o}r}, A., {et~al.} 2016, \apjs, 224, 22,
  \dodoi{10.3847/0067-0049/224/2/22}

\bibitem[{{Lacy} {et~al.}(2007){Lacy}, {Petric}, {Sajina}, {Canalizo},
  {Storrie-Lombardi}, {Armus}, {Fadda}, \& {Marleau}}]{2007AJ....133..186L}
{Lacy}, M., {Petric}, A.~O., {Sajina}, A., {et~al.} 2007, \aj, 133, 186,
  \dodoi{10.1086/509617}

\bibitem[{{Lang} {et~al.}(2016){Lang}, {Hogg}, \&
  {Mykytyn}}]{2016ascl.soft04008L}
{Lang}, D., {Hogg}, D.~W., \& {Mykytyn}, D. 2016, {The Tractor: Probabilistic
  astronomical source detection and measurement}.
\newblock \doeprint{1604.008}

\bibitem[{{Lee} {et~al.}(2021){Lee}, {Lee}, {Lee}, {Suh}, {Cho}, {Byun},
  {Park}, {Herczeg}, {Contreras Pe{\~n}a}, \& {Johnstone}}]{2021Lee}
{Lee}, J.-E., {Lee}, S., {Lee}, S., {et~al.} 2021, \apjl, 916, L20,
  \dodoi{10.3847/2041-8213/ac0d59}

\bibitem[{{Lomb}(1976)}]{1976Lomb}
{Lomb}, N.~R. 1976, \apss, 39, 447, \dodoi{10.1007/BF00648343}

\bibitem[{{Long} {et~al.}(2017){Long}, {Herczeg}, {Pascucci}, {Drabek-Maunder},
  {Mohanty}, {Testi}, {Apai}, {Hendler}, {Henning}, {Manara}, \&
  {Mulders}}]{2017ApJ...844...99L}
{Long}, F., {Herczeg}, G.~J., {Pascucci}, I., {et~al.} 2017, \apj, 844, 99,
  \dodoi{10.3847/1538-4357/aa78fc}

\bibitem[{{Lund} {et~al.}(2017){Lund}, {Silva Aguirre}, {Davies}, {Chaplin},
  {Christensen-Dalsgaard}, {Houdek}, {White}, {Bedding}, {Ball}, {Huber},
  {Antia}, {Lebreton}, {Latham}, {Handberg}, {Verma}, {Basu}, {Casagrande},
  {Justesen}, {Kjeldsen}, \& {Mosumgaard}}]{2017ApJ...835..172L}
{Lund}, M.~N., {Silva Aguirre}, V., {Davies}, G.~R., {et~al.} 2017, \apj, 835,
  172, \dodoi{10.3847/1538-4357/835/2/172}

\bibitem[{{Lynds}(1962)}]{1962ApJS....7....1L}
{Lynds}, B.~T. 1962, \apjs, 7, 1, \dodoi{10.1086/190072}

\bibitem[{{Mainzer} {et~al.}(2011){Mainzer}, {Grav}, {Bauer}, {Masiero},
  {McMillan}, {Cutri}, {Walker}, {Wright}, {Eisenhardt}, {Tholen}, {Spahr},
  {Jedicke}, {Denneau}, {DeBaun}, {Elsbury}, {Gautier}, {Gomillion}, {Hand},
  {Mo}, {Watkins}, {Wilkins}, {Bryngelson}, {Del Pino Molina}, {Desai},
  {G{\'o}mez Camus}, {Hidalgo}, {Konstantopoulos}, {Larsen}, {Maleszewski},
  {Malkan}, {Mauduit}, {Mullan}, {Olszewski}, {Pforr}, {Saro}, {Scotti}, \&
  {Wasserman}}]{2011Mainzer}
{Mainzer}, A., {Grav}, T., {Bauer}, J., {et~al.} 2011, \apj, 743, 156,
  \dodoi{10.1088/0004-637X/743/2/156}

\bibitem[{{Mainzer} {et~al.}(2014){Mainzer}, {Bauer}, {Cutri}, {Grav},
  {Masiero}, {Beck}, {Clarkson}, {Conrow}, {Dailey}, {Eisenhardt}, {Fabinsky},
  {Fajardo-Acosta}, {Fowler}, {Gelino}, {Grillmair}, {Heinrichsen}, {Kendall},
  {Kirkpatrick}, {Liu}, {Masci}, {McCallon}, {Nugent}, {Papin}, {Rice},
  {Royer}, {Ryan}, {Sevilla}, {Sonnett}, {Stevenson}, {Thompson}, {Wheelock},
  {Wiemer}, {Wittman}, {Wright}, \& {Yan}}]{2014Mainzer}
{Mainzer}, A., {Bauer}, J., {Cutri}, R.~M., {et~al.} 2014, \apj, 792, 30,
  \dodoi{10.1088/0004-637X/792/1/30}

\bibitem[{{Majewski} {et~al.}(2011){Majewski}, {Zasowski}, \&
  {Nidever}}]{2011ApJ...739...25M}
{Majewski}, S.~R., {Zasowski}, G., \& {Nidever}, D.~L. 2011, \apj, 739, 25,
  \dodoi{10.1088/0004-637X/739/1/25}

\bibitem[{{Mallick} {et~al.}(2013){Mallick}, {Kumar}, {Ojha}, {Bachiller},
  {Samal}, \& {Pirogov}}]{Mallick.2013}
{Mallick}, K.~K., {Kumar}, M.~S.~N., {Ojha}, D.~K., {et~al.} 2013, \apj, 779,
  113, \dodoi{10.1088/0004-637X/779/2/113}

\bibitem[{{McGuire}(2022)}]{2022ApJS..259...30M}
{McGuire}, B.~A. 2022, \apjs, 259, 30, \dodoi{10.3847/1538-4365/ac2a48}

\bibitem[{{McInnes} {et~al.}(2017){McInnes}, {Healy}, \&
  {Astels}}]{2017JOSS....2..205M}
{McInnes}, L., {Healy}, J., \& {Astels}, S. 2017, The Journal of Open Source
  Software, 2, 205, \dodoi{10.21105/joss.00205}

\bibitem[{{Megeath} {et~al.}(2012){Megeath}, {Gutermuth}, {Muzerolle},
  {Kryukova}, {Flaherty}, {Hora}, {Allen}, {Hartmann}, {Myers}, {Pipher},
  {Stauffer}, {Young}, \& {Fazio}}]{Megeath.2012}
{Megeath}, S.~T., {Gutermuth}, R., {Muzerolle}, J., {et~al.} 2012, \aj, 144,
  192, \dodoi{10.1088/0004-6256/144/6/192}

\bibitem[{{Meingast} \& {Alves}(2019)}]{2019A&A...621L...3M}
{Meingast}, S., \& {Alves}, J. 2019, \aap, 621, L3,
  \dodoi{10.1051/0004-6361/201834622}

\bibitem[{{Meixner} {et~al.}(2006){Meixner}, {Gordon}, {Indebetouw}, {Hora},
  {Whitney}, {Blum}, {Reach}, {Bernard}, {Meade}, {Babler}, {Engelbracht},
  {For}, {Misselt}, {Vijh}, {Leitherer}, {Cohen}, {Churchwell}, {Boulanger},
  {Frogel}, {Fukui}, {Gallagher}, {Gorjian}, {Harris}, {Kelly}, {Kawamura},
  {Kim}, {Latter}, {Madden}, {Markwick-Kemper}, {Mizuno}, {Mizuno}, {Mould},
  {Nota}, {Oey}, {Olsen}, {Onishi}, {Paladini}, {Panagia}, {Perez-Gonzalez},
  {Shibai}, {Sato}, {Smith}, {Staveley-Smith}, {Tielens}, {Ueta}, {van Dyk},
  {Volk}, {Werner}, \& {Zaritsky}}]{2006AJ....132.2268M}
{Meixner}, M., {Gordon}, K.~D., {Indebetouw}, R., {et~al.} 2006, \aj, 132,
  2268, \dodoi{10.1086/508185}

\bibitem[{{Mintz} {et~al.}(2021){Mintz}, {Hora}, \& {Winston}}]{Mintz.2021}
{Mintz}, A., {Hora}, J.~L., \& {Winston}, E. 2021, \aj, 162, 236,
  \dodoi{10.3847/1538-3881/ac2149}

\bibitem[{{Morales-Calder{\'o}n} {et~al.}(2009){Morales-Calder{\'o}n},
  {Stauffer}, {Rebull}, {Whitney}, {Barrado y Navascu{\'e}s}, {Ardila}, {Song},
  {Brooke}, {Hartmann}, \& {Calvet}}]{Morales-Calderon.2009}
{Morales-Calder{\'o}n}, M., {Stauffer}, J.~R., {Rebull}, L., {et~al.} 2009,
  \apj, 702, 1507, \dodoi{10.1088/0004-637X/702/2/1507}

\bibitem[{{Noble} {et~al.}(2013){Noble}, {Fraser}, {Aikawa}, {Pontoppidan}, \&
  {Sakon}}]{2013ApJ...775...85N}
{Noble}, J.~A., {Fraser}, H.~J., {Aikawa}, Y., {Pontoppidan}, K.~M., \&
  {Sakon}, I. 2013, \apj, 775, 85, \dodoi{10.1088/0004-637X/775/2/85}

\bibitem[{{Noble} {et~al.}(2017){Noble}, {Fraser}, {Pontoppidan}, \&
  {Craigon}}]{2017MNRAS.467.4753N}
{Noble}, J.~A., {Fraser}, H.~J., {Pontoppidan}, K.~M., \& {Craigon}, A.~M.
  2017, \mnras, 467, 4753, \dodoi{10.1093/mnras/stx329}

\bibitem[{{Onaka} {et~al.}(2021){Onaka}, {Kimura}, {Sakon}, \&
  {Shimonishi}}]{2021ApJ...916...75O}
{Onaka}, T., {Kimura}, T., {Sakon}, I., \& {Shimonishi}, T. 2021, \apj, 916,
  75, \dodoi{10.3847/1538-4357/ac0531}

\bibitem[{{Park} {et~al.}(2021){Park}, {Lee}, {Contreras Pe{\~n}a},
  {Johnstone}, {Herczeg}, {Lee}, {Lee}, {Bhardwaj}, \&
  {Moriarty-Schieven}}]{2021ApJ...920..132P}
{Park}, W., {Lee}, J.-E., {Contreras Pe{\~n}a}, C., {et~al.} 2021, \apj, 920,
  132, \dodoi{10.3847/1538-4357/ac1745}

\bibitem[{{Pontoppidan} {et~al.}(2010){Pontoppidan}, {Salyk}, {Blake},
  {Meijerink}, {Carr}, \& {Najita}}]{2010ApJ...720..887P}
{Pontoppidan}, K.~M., {Salyk}, C., {Blake}, G.~A., {et~al.} 2010, \apj, 720,
  887, \dodoi{10.1088/0004-637X/720/1/887}

\bibitem[{{Ragan} {et~al.}(2009){Ragan}, {Bergin}, \& {Gutermuth}}]{Ragan.2009}
{Ragan}, S.~E., {Bergin}, E.~A., \& {Gutermuth}, R.~A. 2009, \apj, 698, 324,
  \dodoi{10.1088/0004-637X/698/1/324}

\bibitem[{{Rapson} {et~al.}(2014){Rapson}, {Pipher}, {Gutermuth}, {Megeath},
  {Allen}, {Myers}, \& {Allen}}]{Rapson.2014}
{Rapson}, V.~A., {Pipher}, J.~L., {Gutermuth}, R.~A., {et~al.} 2014, \apj, 794,
  124, \dodoi{10.1088/0004-637X/794/2/124}

\bibitem[{{Rayner} {et~al.}(2003){Rayner}, {Toomey}, {Onaka}, {Denault},
  {Stahlberger}, {Vacca}, {Cushing}, \& {Wang}}]{2003PASP..115..362R}
{Rayner}, J.~T., {Toomey}, D.~W., {Onaka}, P.~M., {et~al.} 2003, \pasp, 115,
  362, \dodoi{10.1086/367745}

\bibitem[{{Reach} {et~al.}(2005){Reach}, {Megeath}, {Cohen}, {Hora}, {Carey},
  {Surace}, {Willner}, {Barmby}, {Wilson}, {Glaccum}, {Lowrance}, {Marengo}, \&
  {Fazio}}]{2005PASP..117..978R}
{Reach}, W.~T., {Megeath}, S.~T., {Cohen}, M., {et~al.} 2005, \pasp, 117, 978,
  \dodoi{10.1086/432670}

\bibitem[{{Rebull}(2015)}]{Rebull.2015}
{Rebull}, L.~M. 2015, \aj, 150, 17, \dodoi{10.1088/0004-6256/150/1/17}

\bibitem[{{Rebull} {et~al.}(2020){Rebull}, {Stauffer}, {Cody}, {Hillenbrand},
  {Bouvier}, {Roggero}, \& {David}}]{Rebull.2020}
{Rebull}, L.~M., {Stauffer}, J.~R., {Cody}, A.~M., {et~al.} 2020, \aj, 159,
  273, \dodoi{10.3847/1538-3881/ab893c}

\bibitem[{{Rebull} {et~al.}(2011){Rebull}, {Guieu}, {Stauffer}, {Hillenbrand},
  {Noriega-Crespo}, {Stapelfeldt}, {Carey}, {Carpenter}, {Cole}, {Padgett},
  {Strom}, \& {Wolff}}]{Rebull.2011}
{Rebull}, L.~M., {Guieu}, S., {Stauffer}, J.~R., {et~al.} 2011, \apjs, 193, 25,
  \dodoi{10.1088/0067-0049/193/2/25}

\bibitem[{{Retes-Romero} {et~al.}(2017){Retes-Romero}, {Mayya}, {Luna}, \&
  {Carrasco}}]{Retes-Romero.2017}
{Retes-Romero}, R., {Mayya}, Y.~D., {Luna}, A., \& {Carrasco}, L. 2017, \apj,
  839, 113, \dodoi{10.3847/1538-4357/aa6afc}

\bibitem[{{Rieke} \& {Lebofsky}(1985)}]{1985ApJ...288..618R}
{Rieke}, G.~H., \& {Lebofsky}, M.~J. 1985, \apj, 288, 618,
  \dodoi{10.1086/162827}

\bibitem[{{Rivera-Ingraham} {et~al.}(2011){Rivera-Ingraham}, {Martin},
  {Polychroni}, \& {Moore}}]{Rivera-Ingraham.2011}
{Rivera-Ingraham}, A., {Martin}, P.~G., {Polychroni}, D., \& {Moore}, T. J.~T.
  2011, \apj, 743, 39, \dodoi{10.1088/0004-637X/743/1/39}

\bibitem[{{Riviere-Marichalar} {et~al.}(2016){Riviere-Marichalar},
  {Mer{\'\i}n}, {Kamp}, {Eiroa}, \& {Montesinos}}]{Riviere-Marichalar.2016}
{Riviere-Marichalar}, P., {Mer{\'\i}n}, B., {Kamp}, I., {Eiroa}, C., \&
  {Montesinos}, B. 2016, \aap, 594, A59, \dodoi{10.1051/0004-6361/201527829}

\bibitem[{{R{\"o}ser} {et~al.}(2019){R{\"o}ser}, {Schilbach}, \&
  {Goldman}}]{2019A&A...621L...2R}
{R{\"o}ser}, S., {Schilbach}, E., \& {Goldman}, B. 2019, \aap, 621, L2,
  \dodoi{10.1051/0004-6361/201834608}

\bibitem[{{Rowles} \& {Froebrich}(2009)}]{2009MNRAS.395.1640R}
{Rowles}, J., \& {Froebrich}, D. 2009, \mnras, 395, 1640,
  \dodoi{10.1111/j.1365-2966.2009.14655.x}

\bibitem[{{Rubio} {et~al.}(1991){Rubio}, {Garay}, {Montani}, \&
  {Thaddeus}}]{1991ApJ...368..173R}
{Rubio}, M., {Garay}, G., {Montani}, J., \& {Thaddeus}, P. 1991, \apj, 368,
  173, \dodoi{10.1086/169680}

\bibitem[{{Saral} {et~al.}(2015){Saral}, {Hora}, {Willis}, {Koenig},
  {Gutermuth}, \& {Saygac}}]{Saral.2015}
{Saral}, G., {Hora}, J.~L., {Willis}, S.~E., {et~al.} 2015, \apj, 813, 25,
  \dodoi{10.1088/0004-637X/813/1/25}

\bibitem[{{Saral} {et~al.}(2017){Saral}, {Hora}, {Audard}, {Koenig},
  {Mart{\'\i}nez-Galarza}, {Motte}, {Nguyen-Luong}, {Saygac}, \&
  {Smith}}]{Saral.2017}
{Saral}, G., {Hora}, J.~L., {Audard}, M., {et~al.} 2017, \apj, 839, 108,
  \dodoi{10.3847/1538-4357/aa6575}

\bibitem[{{Scargle}(1989)}]{1989Scargle}
{Scargle}, J.~D. 1989, \apj, 343, 874, \dodoi{10.1086/167757}

\bibitem[{{Schlafly} {et~al.}(2018){Schlafly}, {Green}, {Lang}, {Daylan},
  {Finkbeiner}, {Lee}, {Meisner}, {Schlegel}, \&
  {Valdes}}]{2018ApJS..234...39S}
{Schlafly}, E.~F., {Green}, G.~M., {Lang}, D., {et~al.} 2018, \apjs, 234, 39,
  \dodoi{10.3847/1538-4365/aaa3e2}

\bibitem[{{Skrutskie} {et~al.}(2006){Skrutskie}, {Cutri}, {Stiening},
  {Weinberg}, {Schneider}, {Carpenter}, {Beichman}, {Capps}, {Chester},
  {Elias}, {Huchra}, {Liebert}, {Lonsdale}, {Monet}, {Price}, {Seitzer},
  {Jarrett}, {Kirkpatrick}, {Gizis}, {Howard}, {Evans}, {Fowler}, {Fullmer},
  {Hurt}, {Light}, {Kopan}, {Marsh}, {McCallon}, {Tam}, {Van Dyk}, \&
  {Wheelock}}]{2006AJ....131.1163S}
{Skrutskie}, M.~F., {Cutri}, R.~M., {Stiening}, R., {et~al.} 2006, \aj, 131,
  1163, \dodoi{10.1086/498708}

\bibitem[{{Stern} {et~al.}(2005){Stern}, {Eisenhardt}, {Gorjian}, {Kochanek},
  {Caldwell}, {Eisenstein}, {Brodwin}, {Brown}, {Cool}, {Dey}, {Green},
  {Jannuzi}, {Murray}, {Pahre}, \& {Willner}}]{2005ApJ...631..163S}
{Stern}, D., {Eisenhardt}, P., {Gorjian}, V., {et~al.} 2005, \apj, 631, 163,
  \dodoi{10.1086/432523}

\bibitem[{{Suh}(2021)}]{2021ApJS..256...43S}
{Suh}, K.-W. 2021, \apjs, 256, 43, \dodoi{10.3847/1538-4365/ac1274}

\bibitem[{{Taylor}(2005)}]{2005ASPC..347...29T}
{Taylor}, M.~B. 2005, in Astronomical Society of the Pacific Conference Series,
  Vol. 347, Astronomical Data Analysis Software and Systems XIV, ed.
  P.~{Shopbell}, M.~{Britton}, \& R.~{Ebert}, 29

\bibitem[{{van Broekhuizen} {et~al.}(2005){van Broekhuizen}, {Pontoppidan},
  {Fraser}, \& {van Dishoeck}}]{2005A&A...441..249V}
{van Broekhuizen}, F.~A., {Pontoppidan}, K.~M., {Fraser}, H.~J., \& {van
  Dishoeck}, E.~F. 2005, \aap, 441, 249, \dodoi{10.1051/0004-6361:20041711}

\bibitem[{{Venuti} {et~al.}(2018){Venuti}, {Prisinzano}, {Sacco}, {Flaccomio},
  {Bonito}, {Damiani}, {Micela}, {Guarcello}, {Randich}, {Stauffer}, {Cody},
  {Jeffries}, {Alencar}, {Alfaro}, {Lanzafame}, {Pancino}, {Bayo}, {Carraro},
  {Costado}, {Frasca}, {Jofr{\'e}}, {Morbidelli}, {Sousa}, \&
  {Zaggia}}]{Venuti.2018}
{Venuti}, L., {Prisinzano}, L., {Sacco}, G.~G., {et~al.} 2018, \aap, 609, A10,
  \dodoi{10.1051/0004-6361/201731103}

\bibitem[{{Watson} {et~al.}(2009){Watson}, {Leisenring}, {Furlan}, {Bohac},
  {Sargent}, {Forrest}, {Calvet}, {Hartmann}, {Nordhaus}, {Green}, {Kim},
  {Sloan}, {Chen}, {Keller}, {d'Alessio}, {Najita}, {Uchida}, \&
  {Houck}}]{2009ApJS..180...84W}
{Watson}, D.~M., {Leisenring}, J.~M., {Furlan}, E., {et~al.} 2009, \apjs, 180,
  84, \dodoi{10.1088/0067-0049/180/1/84}

\bibitem[{{Willis} {et~al.}(2013){Willis}, {Marengo}, {Allen}, {Fazio},
  {Smith}, \& {Carey}}]{Willis.2013}
{Willis}, S., {Marengo}, M., {Allen}, L., {et~al.} 2013, \apj, 778, 96,
  \dodoi{10.1088/0004-637X/778/2/96}

\bibitem[{{Winston} {et~al.}(2019){Winston}, {Hora}, {Gutermuth}, \&
  {Tolls}}]{2019ApJ...880....9W}
{Winston}, E., {Hora}, J., {Gutermuth}, R., \& {Tolls}, V. 2019, \apj, 880, 9,
  \dodoi{10.3847/1538-4357/ab27c8}

\bibitem[{{Winston} {et~al.}(2020){Winston}, {Hora}, \&
  {Tolls}}]{2020AJ....160...68W}
{Winston}, E., {Hora}, J.~L., \& {Tolls}, V. 2020, \aj, 160, 68,
  \dodoi{10.3847/1538-3881/ab99c8}

\bibitem[{{Wolk} {et~al.}(2015){Wolk}, {G{\"u}nther}, {Poppenhaeger}, {Cody},
  {Rebull}, {Forbrich}, {Gutermuth}, {Hillenbrand}, {Plavchan}, {Stauffer},
  {Covey}, \& {Song}}]{Wolk.2015}
{Wolk}, S.~J., {G{\"u}nther}, H.~M., {Poppenhaeger}, K., {et~al.} 2015, \aj,
  150, 145, \dodoi{10.1088/0004-6256/150/5/145}

\bibitem[{{Wright} {et~al.}(2010){Wright}, {Eisenhardt}, {Mainzer}, {Ressler},
  {Cutri}, {Jarrett}, {Kirkpatrick}, {Padgett}, {McMillan}, {Skrutskie},
  {Stanford}, {Cohen}, {Walker}, {Mather}, {Leisawitz}, {Gautier}, {McLean},
  {Benford}, {Lonsdale}, {Blain}, {Mendez}, {Irace}, {Duval}, {Liu}, {Royer},
  {Heinrichsen}, {Howard}, {Shannon}, {Kendall}, {Walsh}, {Larsen}, {Cardon},
  {Schick}, {Schwalm}, {Abid}, {Fabinsky}, {Naes}, \&
  {Tsai}}]{2010AJ....140.1868W}
{Wright}, E.~L., {Eisenhardt}, P. R.~M., {Mainzer}, A.~K., {et~al.} 2010, \aj,
  140, 1868, \dodoi{10.1088/0004-6256/140/6/1868}

\bibitem[{{Zasowski} {et~al.}(2009){Zasowski}, {Majewski}, {Indebetouw},
  {Meade}, {Nidever}, {Patterson}, {Babler}, {Skrutskie}, {Watson}, {Whitney},
  \& {Churchwell}}]{2009ApJ...707..510Z}
{Zasowski}, G., {Majewski}, S.~R., {Indebetouw}, R., {et~al.} 2009, \apj, 707,
  510, \dodoi{10.1088/0004-637X/707/1/510}

\bibitem[{{Zucker} {et~al.}(2019){Zucker}, {Speagle}, {Schlafly}, {Green},
  {Finkbeiner}, {Goodman}, \& {Alves}}]{2019ApJ...879..125Z}
{Zucker}, C., {Speagle}, J.~S., {Schlafly}, E.~F., {et~al.} 2019, \apj, 879,
  125, \dodoi{10.3847/1538-4357/ab2388}

\end{thebibliography}
		\bibliographystyle{aasjournal}

\end{document}